\documentclass[reprint, superscriptaddress, amsmath, amssymb, aps, prx]{revtex4-2}

\usepackage{graphicx}
\usepackage{dcolumn}
\usepackage{bm}
\usepackage[english]{babel}
\usepackage{gensymb}
\usepackage{siunitx}
\usepackage[normalem]{ulem}
\usepackage{color}
\usepackage{multirow}
\usepackage{array}
\usepackage{tabularx}
\usepackage{amsmath}
\usepackage{amssymb}
\usepackage{pifont}
\usepackage{xr-hyper} 
\usepackage{hyperref}
\hypersetup{
}
\usepackage{cleveref}
\usepackage{subfigure}
\usepackage[version=4]{mhchem}
\usepackage{booktabs}

\usepackage{chemformula} 
\usepackage[T1]{fontenc} 
\usepackage{xcolor}
\usepackage{gensymb}
\usepackage{yfonts}
\usepackage[compat=1.1.0]{tikz-feynman}
\usepackage{calligra}
\usepackage[T1]{fontenc}

\usepackage{braket}

\makeatletter
\renewcommand*\env@matrix[1][*\c@MaxMatrixCols c]{%
  \hskip -\arraycolsep
  \let\@ifnextchar\new@ifnextchar
  \array{#1}}
\makeatother

\crefname{figure}{Fig.}{Figs.}
\Crefname{figure}{Figure}{Figures}

\crefname{table}{Table}{Tables}
\Crefname{table}{Table}{Tables}

\crefname{equation}{Eq.}{Eqs.}
\Crefname{equation}{Equation}{Equations}

\crefname{section}{Sec.}{Secs.}
\Crefname{section}{Section}{Sections}

\crefname{appendix}{Appendix}{Appendices}
\Crefname{appendix}{Appendix}{Appendices}

\DeclareMathAlphabet{\mathcal}{OMS}{cmsy}{m}{n}

\newcommand{\kk}{\mathbf{k}}
\begin{document}

\title[]{Memory-dependent electronic friction for nonadiabatic dynamics at metal surfaces}

\author{Xuexun Lu}
\affiliation{Department of Chemistry, University of Warwick, Coventry, CV4 7AL, United Kingdom}

\author{Connor L. Box}
\affiliation{Yusuf Hamied Department of Chemistry, University of Cambridge, Cambridge, CB2 1EW, United Kingdom}

\author{Nils Hertl}
\affiliation{Department of Chemistry, University of Warwick, Coventry, CV4 7AL, United Kingdom}
\affiliation{Department of Physics, University of Warwick, Coventry, CV4 7AL, United Kingdom}

\author{Reinhard J. Maurer}
\email{reinhard.maurer@univie.ac.at}
\affiliation{Department of Chemistry, University of Warwick, Coventry, CV4 7AL, United Kingdom}
\affiliation{University of Vienna, Faculty of Physics, University of Vienna, Kolingasse 14{--}16, 1090 Vienna, Austria}
\affiliation{Institute of Physical Chemistry, Georg-August University, Göttingen 37077, Germany}
\affiliation{Max-Planck-Institute for Multidisciplinary Sciences, Göttingen 37077, Germany}


\begin{abstract}
Electronic excitation induced by nuclear motion is a key energy dissipation channel in chemical dynamics at metal surfaces. Here, nonadiabatic effects can be treated via molecular dynamics with electronic friction, where they act as frictional drag and fluctuation force contributions. Commonly, the Markov approximation is imposed, so memory effects are ignored. A theoretical formalism is presented to evaluate tensorial and configuration-dependent electronic friction memory kernels from first principles. We evaluate friction kernels for Newns--Anderson Hamiltonian models as well as within Kohn--Sham density functional theory and analyse their mathematical properties and configuration dependence. For hyperthermal atomic and diatomic scattering, memory effects arising from frequency and configuration dependence of electronic friction affect energy exchange between adsorbate and metal electrons. Memory effects lead to an increase of vibrational and a reduction of translational   energy loss in the case of nitric oxide scattering on Au(111), leading to an increase of directional anisotropy of friction. Importantly, memory-dependent evaluation of electronic friction removes the need to define a single effective Markovian friction coefficient from the structured frequency-dependent electronic response.
\end{abstract}

\maketitle

\clearpage


\section{Introduction}

Nonadiabatic energy exchange between adsorbates and the electrons of a metal surface measurably affects experimental observables in gas--surface dynamics \cite{saalfrank2006quantum, auerbach21}. Seminal experiments demonstrated nonadiabatic chemicurrents in gas-surface scattering~\cite{gergen2001}, activated diffusion during inelastic electron tunelling spectroscopy of hydrogen (H) on copper~\cite{lamont1995, lauhon2000}, and vibrationally and translationally inelastic scattering of nitric oxide (NO)~\cite{huang2000, kruger2016vibrational}, carbon monoxide (CO)~\cite{schafer2013, wagner2017vibrational} and H atoms from metal surfaces~\cite{buenermann15,Dorenkamp18, hertl22b}.

When the electronic excitations can be treated as a weakly coupled bath, their effect on the nuclei can be described using molecular dynamics with electronic friction (MDEF)~\cite{head-gordon95, dou17, dou18}. Within MDEF, the electrons are treated implicitly as a bath, and the dynamics are entirely governed by the ground-state PES, whilst the coupling between electrons and the nuclei is described by a friction and a random force, resulting in a generalised Langevin equation of motion (GLE). Mixed quantum-classical formalisms of MDEF were independently formulated by Persson and Hellsing~\cite{Persson1982}, Head-Gordon and Tully~\cite{head-gordon95} and Brandbyge \textit{et al.}~\cite{brandbyge_1995_electronically} to study nonadiabatic vibrational relaxation and high-dimensional nonadiabatic dynamics. Initial formalisms considered 0~K and finite-temperature cases for non-interacting electron systems, whereas further works also considered electron-electron interactions~\cite{dou17}. Smith and Hynes introduced electronic friction in the context of electron transfer kinetics in electrochemistry~\cite{smith_electronic_1993}. Electronic friction theory has also been extended to open, non-equilibrium systems~\cite{Luu2012, bode_scattering_2011}, where electronic friction forces coexist with current-induced forces. While electronic friction tensors in closed systems have to be positive semi-definite, in open systems, negative friction can arise. Friction and current-induced forces inherently arise from the viscosity and the stopping power of electrons in a conductor~\cite{nazarov_time-dependent_2005, nazarov_viscous_2024, nazarov_stopping_nodate}.  Recently, for non-equilibrium systems, a direct connection between negative electronic friction and non-Markovianity was identified~\cite{preston2026}. A full quantum dynamic treatment of electronic friction has also been formulated~\cite{martinazzo_quantum_2022, martinazzo_quantum_2022-1}. Electronic friction and its limitations have been extensively studied with the help of parametrised Hamiltonian models~\cite{dou2016a, gardner2023, gardner2023a}.

The friction kernel in the GLE is generally time-dependent as it relates to the time-time correlation function of nonadiabatic coupling along a classical dynamical trajectory. Associated with this are memory effects that conceptually arise from the coarse-graining of electronic degrees of freedom into a bath with a broad spectrum of frequencies \cite{head-gordon95,Zwanzig2001}, which are coupled to the nuclear motion. The resulting non-Markovian system–bath dynamics include the dependence of the system-bath coupling on the configuration of the system, but also the variation of system-bath coupling with the excitation frequency of the electrons. 

Although strategies to incorporate memory effects into electronic friction simulations have been proposed \cite{olsen2010, trenins2025}, in practical gas-surface simulations, MDEF has so far been deployed in the Markov approximation, i.e., memory effects in the friction force are omitted. Hence, the friction force reduces to a product of a Markovian (memoryless) friction tensor and the velocity vector of the nuclei. This approach has been employed in combination with first-principles electronic structure evaluation of electronic friction, for example through the stopping power associated with scattering of ions in the homogeneous electron gas (the so-called local density friction approximation)~\cite{li_wahnstrom92b, juaristi08, gerrits20} or by direct evaluation of the electron-phonon linear response in Kohn-Sham density functional theory (DFT) \cite{hellsing1984,  maurer16, box_ab_2023}. 
Markovian MDEF based on \textit{ab initio} data (directly or through surrogate models) has been successfully applied to simulate the translational energy loss in H atom scattering experiments from metal surfaces \cite{Dorenkamp18, hertl22b, box_room_2024}, laser-induced desorption \cite{fuchsel11, zugec2024understanding, spears_role_2026}, and electron-hole-pair-driven vibrational relaxation rates \cite{Persson1982, maurer16, hertl2026}. In the case of the vibrationally inelastic scattering of NO from Au(111), it was shown that Markovian MDEF significantly underestimates vibrational energy loss compared to experiment for highly excited initial vibrational states~\cite{box_2020_determining}. When considering the average increase of friction due to the coupling of the molecule with higher-lying excited states, Box \textit{et al.}~\cite{box_2020_determining} were able to reach better agreement with experiment, suggesting that inclusion of memory effects may be able to extend the applicability regime of electronic friction.

In this work, we introduce a practically feasible approach to evaluate memory-dependent electronic friction kernels from first principles. Building on previously established mixed quantum-classical formalisms \cite{head-gordon95, brandbyge_1995_electronically, Luu2012}, we provide a consistent derivation based on 
a local linear-response formulation in which the frequency-dependent friction kernel can be evaluated from electronic-structure quantities at a single nuclear configuration.
This is an important prerequisite to enable the numerically robust calculation of the friction kernel from first principles. We investigate the mathematical and numerical properties of the memory-dependent friction kernel for two parametrised Newns-Anderson model Hamiltonians that represent atom \cite{erpenbeck2018} and molecule scattering \cite{gardner2023a}. Furthermore, we analyse the high-dimensional friction spectrum of \textit{ab initio} friction tensors for NO on a gold surface calculated from Kohn-Sham DFT. For the studied model systems, memory effects can lead to an increase or a reduction of energy dissipation, depending on the configurations that are dynamically sampled. At state crossings, electronic friction memory kernels reflect overdamped dynamics and the inclusion of memory reduces energy dissipation compared to the Markovian limit. At all other configurations, memory kernels reflect underdamped dynamics where the electron-phonon resonance point sits at a frequency, $\omega>0$. Using the classical path approximation, we quantify the effect of memory on nonadiabatic energy dissipation during scattering dynamics. In NO on Au(111), memory effects enhance vibrational energy loss and reduce translational energy loss, further emphasising the anisotropic nature of electronic friction. We compare against two types of Markovian approximations of the friction kernel; neither can reproduce the non-Markovian behaviour.  Our findings thus have important implications for previous MDEF studies \cite{box_2020_determining, meng2024, hong2025, bombin2026}, as Markovian MDEF treatment may significantly affect nonadiabatic energy dissipation in hyperthermal scattering. 

\section{Theory}

\subsection{Molecular dynamics with electronic friction}\label{sec:MDEF}
Molecular dynamics with electronic friction describes nuclear motion on a ground state potential energy surface, $E_0$, under the influence of weak nonadiabatic coupling with low-lying electronic excited states in a metal. The equation of motion is a nonlinear generalised Langevin equation~\cite{head-gordon95}: 
\begin{equation}
\label{eq:MDEF-g}
    \mathbf{M} \ddot{\mathbf{x}} = -\nabla_{\mathbf{x}} E_0 - \int_{0}^{t} \mathrm{d}t' \, \boldsymbol{\mathcal{K}}(t-t'\,;\mathbf{x}(t), \mathbf{x}(t')) \, \dot{\mathbf{x}}(t') + \boldsymbol{\xi}^{\mathrm{c}}(t),
\end{equation}
where $\mathbf{M}$ is the mass matrix and $\mathbf{x}$ is the nuclear position vector. The explicit propagation of the electronic degrees of freedom is integrated out and represented as a frictional drag term with a non-Markovian friction kernel,  $\mathcal{K}$, that is symmetric in the coordinate indices and a corresponding coloured noise term, $\boldsymbol{\xi}^c$.

For individual modes, $\mu$, this becomes
\begin{equation}
\label{eq:MDEF-kmode}
\begin{split}
m_\mu \ddot{x}_\mu
&= -\frac{\partial E_0}{\partial x_\mu}
   - \sum_\nu \int_0^t \mathrm{d}t'\,
   \mathcal{K}_{\mu\nu}
   \bigl(
      t-t'\,;
      \mathbf{x}(t),
      \mathbf{x}(t')
   \bigr)
   \dot{x}_\nu(t') \\
&\qquad
   + \xi_\mu^{\mathrm{c}}(t).
\end{split}
\end{equation}
To retain a thermal temperature for the system, the fluctuation–dissipation theorem has to be satisfied: 
\begin{equation}
    \label{eq:fdt}
    \langle \xi_{\mu}^{\mathrm{c}}(t)\,
            \xi_{\nu}^{\mathrm{c}}(t') \rangle 
    = k_{\mathrm{B}}T\,
      \mathcal{K}_{\mu\nu}\bigl(
         t-t'\,;
         \mathbf{x}(t),
         \mathbf{x}(t')
      \bigr)
\end{equation}
with $k_{\mathrm{B}}$ and $T$ representing the Boltzmann constant and temperature,  respectively.

Head-Gordon and Tully \cite{head-gordon95} remove the explicit dependence on electronic degrees of freedom by working in a gauge-fixed representation within the adiabatic ground state $|0\rangle$ and truncating at first order in the nonadiabatic coupling and the electronic deviations from $|0\rangle$. Each electronic mode then behaves as an independent harmonic oscillator linearly coupled to the nuclear velocity $\dot{\mathbf{x}}$. The resulting memory kernel $\mathcal{K}_{\mu\nu}$ captures the retarded response of the electronic subsystem to nuclear motion, incorporating the nonadiabatic coupling matrix elements between $\lvert 0 \rangle$ and the continuum of many-body excited states
$\lvert n \rangle$ accessible via low-energy electron--hole pair
excitations near the Fermi level.
Expressed in the normal-mode representation, this kernel reads:
\begin{equation}
\label{eq:kernel-microscopic}
\begin{split}
\mathcal{K}_{\mu\nu}
    \bigl(
        t-t'\,;
        \mathbf{x}(t),
        \mathbf{x}(t')
    \bigr)
&= 2\hbar \sum_n
    d_{n0}^{\mu}\!\bigl(\mathbf{x}(t)\bigr)\,
    d_{n0}^{\nu}\!\bigl(\mathbf{x}(t')\bigr) \\
&\qquad\times
    \omega_{n0}\,
    \cos\!\bigl[\omega_{n0}(t-t')\bigr].
\end{split}
\end{equation}
The excitation frequency $\omega_{n0} = (E_n - E_0)/\hbar$ of state $\lvert n \rangle$ is treated as constant over the memory window ~\cite{head-gordon95}. The nonadiabatic coupling between the ground state $\lvert 0 \rangle$ and excited state
$\lvert n \rangle$ at mode~$\mu$ is
\begin{equation}
\label{eq:nac}
  d_{n0}^{\mu}(\mathbf{x})
  = \bigl\langle n \big\rvert
    \frac{\partial}{\partial x_\mu}
    \big\lvert 0 \bigr\rangle.
\end{equation}
 The product $d_{n0}^{\mu}(\mathbf{x}(t))\,d_{n0}^{\nu}(\mathbf{x}(t'))$ in \cref{eq:kernel-microscopic} is the origin of the explicit dependence of the kernel on both
$\mathbf{x}(t)$ and $\mathbf{x}(t')$. 

For metals, the electron--hole pair excitation
spectrum forms a continuum near the Fermi level and the discrete sum over many-body excited states in \cref{eq:kernel-microscopic}
can be written as a continuum integral by introducing the spectral density tensor 
\begin{equation}
\label{eq:spectral-density}
  \mathcal{J}_{\mu\nu}(\omega\,;\mathbf{x}(t),\mathbf{x}(t'))
  = 2\hbar \sum_n
    d^{\mu}_{n0}(\mathbf{x}(t))\,
    d^{\nu}_{n0}(\mathbf{x}(t'))\,
    \delta(\omega - \omega_{n0}),
\end{equation}
which encodes the coupling-weighted density of electron--hole pair
excitations at frequency~$\omega$, resolved by nuclear mode pair
$(\mu,\nu)$ and evaluated at nuclear configurations. The memory kernel then takes the compact form
\begin{equation}
\label{eq:kernel-continuum}
\begin{split}
  \mathcal{K}_{\mu\nu}\bigl( t-t'\,;\mathbf{x}(t),\mathbf{x}(t') \bigr)
  = {} & \int_0^{\infty} \mathrm{d}\omega\; \mathcal{J}_{\mu\nu}\bigl(\omega\,;\mathbf{x}(t),\mathbf{x}(t') \bigr) \\
  & \times \omega\,\cos\!\bigl[\omega(t-t')\bigr].
\end{split}
\end{equation}
The passage from \cref{eq:kernel-microscopic} to
\cref{eq:kernel-continuum} is exact.

If the spectrum
$\omega\mathcal{J}_{\mu\nu}(\omega;\,\mathbf{x},\mathbf{x}')$ varies negligibly
over the frequency scale of nuclear motion, known as the constant coupling limit, the cosine transform reduces to a Dirac delta distribution $\delta(t-t')$, such that the memory
kernel becomes local in time and reduces to a Markovian friction tensor,  $\eta_{\mu\nu}(\mathbf{x}(t))$:
\begin{equation}
\label{eq:markov-limit}
  \mathcal{K}_{\mu\nu}\!\bigl(t-t'\,;\mathbf{x}(t),\mathbf{x}(t')\bigr)
  \;\longrightarrow\;
  \eta_{\mu\nu}(\mathbf{x}(t))\,\delta(t-t'),
\end{equation}
In this limit, the generalised Langevin equation (\cref{eq:MDEF-kmode}) reduces to the commonly applied Markovian form of MDEF: 
\begin{equation}
\label{eq:MDEF-markov}
  m_\mu \ddot{x}_\mu
  = -\frac{\partial E_0}{\partial x_\mu}
    - \sum_{\nu} \eta_{\mu\nu}(\mathbf{x})\,\dot{x}_\nu
    + \xi_\mu^{\mathrm{w}}(t),
\end{equation}
where the fluctuation term $\xi_\mu^{\mathrm{w}}$ is now white noise, satisfying
$\langle \xi_\mu^{\mathrm{w}}(t)\,\xi_\nu^{\mathrm{w}}(t') \rangle
= 2\,k_{\mathrm{B}}T\,\eta_{\mu \nu}(\mathbf{x})\,\delta(t-t')$.
    
\subsection{Electronic friction in the linear-coupling regime}\label{sec:linearisationEHamiltonian}

In the weak electron-phonon coupling regime, the electronic Hamiltonian admits a first-order Taylor expansion with respect to the nuclear displacement about the equilibrium configuration $\tilde{\textbf{x}}_0$:
\begin{equation}
    \label{eq:linear regime}
    \hat{H}_{\text{el}}(\mathbf{x}) \approx \hat{H}_{\text{el}}(\tilde{\mathbf{x}}_0) 
    + \sum_\mu (x_\mu - \tilde{x}_{0,\mu})\,
    \hat{\mathcal{M}}^{\mu}(\tilde{\mathbf{x}}_0)
\end{equation}
where the electron-nuclear coupling operator
\begin{equation}
    \hat{\mathcal{M}}^\mu \equiv \left. 
    \frac{\partial \hat{H}_{\text{el}}(\mathbf{x})}{\partial x_\mu} 
    \right|_{\mathbf{x}=\tilde{\mathbf{x}}_0} 
\end{equation}
represents the response of $\hat{H}_{\text{el}}$ to displacement of the $\mu$-th nuclear coordinate. 

We now restrict ourselves to non-interacting electrons, so that $\hat{H}_{\text{el}}(\mathbf{x})$ is quadratic in the fermion operators. In a single-particle basis $\{|k \rangle\}$, the coupling operator is represented by the Hermitian matrix
\begin{equation}
    \label{eq:Mmatrix}
    \mathcal{M}^\mu_{kl} \;\equiv\; 
    \langle k |\,\hat{\mathcal{M}}^\mu\,| l\rangle.
\end{equation}
We refer to $\mathcal{M}^\mu$ as the electron-nuclear coupling matrix and $\hat{\mathcal{M}}^\mu$ as the conjugate many-body operator. At the single-electron level, it is related to nonadiabatic coupling by
\begin{equation}
    d^{\mu}_{kl} = \frac{\mathcal{M}^\mu_{kl}}{\epsilon_l - \epsilon_k}, 
    \quad (k \neq l), \qquad 
    \hat{H}_{\text{el}}\,|k\rangle = \epsilon_i\,|k\rangle.
\end{equation}

To investigate electronic memory effects during non-equilibrium nuclear dynamics, we adopt the quasi-adiabatic limit discussed by Lü \textit{et al.}~\cite{Luu2012}: the nuclear velocity $\dot{\mathbf{x}}$ is considered relatively small to the electronic correlation rate, so that the perturbation of the nuclei within the memory window is negligible. Hence, all the electronic-structure quantities entering the friction kernel are determined by the instantaneous electronic Hamiltonian $\hat{H}_{\text{el}}(\mathbf{x}(t))$ at the current nuclear configuration, $\textbf{x}(t)$. The equilibrium reference, $\tilde{\textbf{x}}_0$, in \cref{eq:linear regime} is thereby represented as the instantaneous trajectory point,
\begin{equation}
\label{eq:reanchoring}
    \tilde{\textbf{x}}_0 \rightarrow \mathbf{x}(t).
\end{equation}

By applying the Feynman-Vernon influence-functional or Keldysh nonequilibrium Green's function (NEGF) framework~\cite{Luu2012,brandbyge_1995_electronically,FEYNMAN1963118} to the linearized Hamiltonian \cref{eq:linear regime} in the quasi-adiabatic limit, the electronic degrees of freedom are linearly coupled to the nuclear displacements by the operators $\{\hat{\mathcal{M}}^{\mu}(\mathbf{x}(t))\}$. Integrating out the resulting system-bath effects yields a generalised Langevin equation for the nuclei,
\begin{equation}
    \label{eq:GLE-single-position}
\begin{split}
m_\mu \ddot{x}_\mu
&= -\frac{\partial E_0}{\partial x_\mu}
   - \sum_\nu \int_0^t \mathrm{d}t'\,
   \mathcal{K}_{\mu\nu}
   \bigl(
      t-t';\,\mathbf{x}(t)
   \bigr)
   \dot{x}_\nu(t') \\
&\qquad
   + \xi_\mu^{\mathrm{c}}(t)
\end{split}
\end{equation}
with the same fluctuation-dissipation relation as \cref{eq:fdt}, however, the memory kernel $$\mathcal{K}_{\mu \nu}
\bigl(
      t-t';
      \mathbf{x}(t)
\bigr)$$ now depends only on the instantaneous nuclear configuration at the current time $t$, even though it couples the current force to the velocity in the past $t'<t$. This single-position ansatz follows directly from the linearisation in Eq.~\eqref{eq:linear regime}: the coupling operators $\hat{\mathcal{M}}^{\mu}(\tilde{\textbf{x}}_0)$ depend on a single reference configuration, and the full trajectory information enters only through the displacements $(x_\mu-\tilde{x}_{0,\mu})$. This ansatz contrasts with nonlinear formulations such as the one by Head-Gordon and Tully shown in the previous section, in which the coupling at each trajectory time point retains its full configurational dependence and the friction kernel exhibits a two-time structure, $\mathcal{K}_{\mu \nu}
   \bigl(
      t-t'\,;
      \mathbf{x}(t),
      \mathbf{x}(t')
   \bigr)$.
   
A direct first-principles evaluation of the two-time kernel using \textit{ab initio} electronic-structure methods  would require establishing a consistent correspondence between electronic states obtained from separate electronic structure calculations at different nuclear configurations, including robust state tracking through crossings or near-degeneracies and a consistent electronic gauge (phase) along the trajectory. 

Consequently, friction kernels for electronic Hamiltonian models and full \textit{ab initio} calculations based on Kohn-Sham DFT will subsequently be formulated within the local linear-coupling approximation, in which the electron--nuclear coupling matrices are evaluated at a single reference configuration, while the non-Markovian character of the dynamics is retained through the frequency dependence of the resulting friction kernel.

\subsection{Local position-dependent friction kernel}\label{sec:Single position based friction kernel}

Within the electronic Hamiltonian linearisation regime discussed previously, the local position based memory electronic friction kernel for mode $\mu$ and $\nu$ is illustrated as
\begin{equation}
    \label{eq:etat}
    \mathcal{K}_{\mu\nu}(t-t'\,;\mathbf{x}(t)) := \Lambda_{\mu\nu}(t-t'\,;\mathbf{x}(t)) \theta(t-t')
\end{equation}
where $\Lambda_{\mu\nu}(t-t'\,;\mathbf{x}(t))$ represents a symmetrized-in-time correlation function satisfying
$\Lambda_{\mu\nu}(t'-t\,;\mathbf{x}(t)) = \Lambda_{\mu\nu}(t-t'\,;\mathbf{x}(t))$ 
and $\theta$ is the Heaviside step function that ensures causality. For brevity, we suppress the dependence on the instantaneous configuration
$$
\mathbf{x} \equiv \mathbf{x}(t).
$$
All quantities in either frequency space $\omega$ or time space within this section are understood to be evaluated at, and to retain their dependence on, configuration $\mathbf{x}(t)$.

Following the NEGF formalism \cite{Luu2012}, the friction kernel in frequency domain, $\mathcal{K}_{\mu\nu}(\omega)$, is expressed as:
    \begin{equation}
    \label{eq:self-energyfriction}
    \mathcal{K}_{\mu\nu}(\omega\,;\mathbf{x}) := \Lambda_{\mu\nu}(\omega\,;\mathbf{x})/2 = - \frac{\pi}{2\omega} \text{Re} \Gamma_{\mu\nu}(\omega\,;\mathbf{x}) .
    \end{equation}
 The spectral dissipation function $\Lambda_{\mu\nu}$ is determined by $\Gamma_{\mu\nu}$, the interaction-weighted electron-phonon density of states (for spinless electrons). The real part of the latter is related to the retarded electron–hole self-energy components $\Pi^{r}_{\mu\nu}$ and $\Pi^{r}_{\nu\mu}$ through
\begin{equation}
    \label{eq:ReGammaLu}
    \operatorname{Re} \left[ \Gamma_{\mu\nu} \right] = \frac{1}{2\pi} \left[ \operatorname{Im} \Pi_{\mu\nu}^{r} + \operatorname{Im} \Pi_{\nu\mu}^{r} \right].
\end{equation}
The retarded and advanced electron–hole pair self-energies are introduced in the time domain, 
    \begin{equation}
    \label{eq:_self_energy}
    \begin{split}
    \Pi_{\mu\nu}^{r,a}&(t-t'\,;\mathbf{x}) = {} \\
    & -i\hbar \, \operatorname{Tr}
    \left[\mathcal{M}^\mu G_0^{r,a}(t-t'\,;\mathbf{x}) \mathcal{M}^\nu G_0^<(t'-t\,;\mathbf{x})\right] \\
    & -i\hbar \, \operatorname{Tr}
    \left[\mathcal{M}^\mu G_0^<(t-t'\,;\mathbf{x})
    \mathcal{M}^\nu G_0^{a,r}(t'-t\,;\mathbf{x})\right].
    \end{split}
    \end{equation}
Here, $G_0^{r}$, $G_0^{a}$, and $G_0^<$ denote the retarded, advanced, and lesser Green’s functions of the non-interacting electronic system, respectively. And $\mathcal{M}^\mu$ stands for the electron-phonon coupling of mode $\mu$.

Converting \cref{eq:_self_energy} to frequency domain by Fourier transform, we obtain its imaginary part as
    \begin{equation}\label{eq:imaginaryPi_matrix}
    \begin{split}
    \operatorname{Im}\Pi^r_{\mu\nu}(\omega\,;\mathbf{x}) = {} &
    \int \frac{d\epsilon}{2\pi}\;
    \operatorname{Tr}\left[\mathcal{M}^\mu\,\mathcal{A}
    (\epsilon+\hbar\omega\,;\mathbf{x})\, \mathcal{M}^\nu\,
    \mathcal{A}(\epsilon\,;\mathbf{x})\right] \\
    & \times \left[n_F(\epsilon+\hbar\omega)
    - n_F(\epsilon)\right],
    \end{split}
    \end{equation}
    where $\mathcal{A}$ represents the impurity spectral function. The detail of the Green's functions and derivation of \cref{eq:imaginaryPi_matrix} are provided in Appendix\,\ref{ap:retarded-electron-hole-pair-self-energy}.

\subsection{Memory friction kernel for Newns--Anderson Hamiltonian}\label{sec:NAH}
The Newns--Anderson Hamiltonian (NAH) represents a canonical framework for describing the nonadiabatic dynamics of an impurity state coupled to an electronic continuum:

\begin{equation}
\hat{H}_{\text{NA}}(\hat{\mathbf{x}}, \mathbf{\hat{p}}) = \sum_{\mu} \frac{\hat{p}_{\mu}^2}{2m_{\mu}} + U_0(\hat{\mathbf{x}}) + \hat{H}_{\text{el}}(\hat{\mathbf{x}})
\end{equation}

where $\hat{\mathbf{x}}$ and $\mathbf{\hat{p}}$ represent the vectors of nuclear coordinate operators and their conjugate momenta, respectively, for particles with masses $m_\mu$. The index $\mu$ runs over the nuclear degrees of freedom. The potential energy is governed by the state-independent potential $U_0$ and the electronic Hamiltonian $\hat{H}_{\text{el}}$, the latter of which depends parametrically on the nuclear configuration. For a single impurity level coupled to a manifold of reservoir states, the electronic Hamiltonian is
\begin{equation}
\begin{split}
\hat{H}_{\text{el}}(\hat{\mathbf{x}}) = {} &
h(\hat{\mathbf{x}})\hat{d}^{\dagger}\hat{d} 
+ \sum_{k} \epsilon_k\hat{c}_k^{\dagger}\hat{c}_k \\
& + \sum_{k} V_k(\hat{\mathbf{x}})
\left(\hat{d}^{\dagger}\hat{c}_k 
+ \hat{c}_k^{\dagger}\hat{d}\right).
\end{split}
\end{equation}
Here, $\hat{d}^{\dagger}(\hat{d}\,)$ and $\hat{c}_k^{\dagger}(\hat{c}_k)$ are the creation (annihilation) operators for the impurity state and the $k$-th  reservoir state, respectively. The term $h(\hat{\mathbf{x}}) = U_1(\hat{\mathbf{x}}) - U_0(\hat{\mathbf{x}})$ represents the energy of the impurity level relative to the reservoir, effectively defining the difference between the diabatic potential energy surfaces associated with the occupied ($U_1$) and unoccupied ($U_0$) impurity states. And the terms $V_k(\hat{\mathbf{x}})$ describe the coupling between the impurity level to a $k$-th discrete state in the electronic reservoir.

To describe the electronic population transfer between the molecule and the reservoir, we introduce the hybridization function
\begin{equation}
\label{eq:general-hybridistaion}
\Delta(\epsilon\,; \hat{\mathbf{x}}) =
\pi \sum_{k} |V_k(\hat{\mathbf{x}})|^2
\delta(\epsilon - \epsilon_k).
\end{equation}
Under the constant-coupling approximation, the impurity couples to each reservoir state with the same coupling magnitude,
$V_k(\hat{\mathbf{x}})=V(\hat{\mathbf{x}})$ for all $k$. The hybridization function then reduces to
\begin{equation}
\label{eq:constant-hybridistaion}
\Delta(\epsilon\,; \hat{\mathbf{x}}) =
\pi |V(\hat{\mathbf{x}})|^2 \rho(\epsilon),
\end{equation}
where $\rho(\epsilon)$ is the reservoir density of states. Thus, the hybridization function factorizes into the configuration-dependent coupling strength, $|V(\hat{\mathbf{x}})|^2$, and the energy-dependent reservoir density of states, $\rho(\epsilon)$.

In the single impurity limit of the NAH, the orbital trace in $\operatorname{Im}\Pi^r_{\mu\nu}$ (\cref{eq:imaginaryPi_matrix}) collapses to a scalar product. The spectral-function matrix $\mathcal{A}$ accordingly reduces to the scalar impurity spectral function 
\begin{equation}
    \label{eq:Ja}
    \mathcal{A}(\epsilon\,; \hat{\mathbf{x}}) = \frac{2 \Delta\left(\epsilon\,; \hat{\mathbf{x}}\right)}{\left[\epsilon-h(\hat{\mathbf{x}})-\mathcal{H}\left(\epsilon\,; \hat{\mathbf{x}}\right)\right]^2+\Delta\left(\epsilon\,; \hat{\mathbf{x}}\right)^2},
\end{equation}
where the energy shift $\mathcal{H}$ is the Hilbert transform of the hybridization function: 
\begin{equation}
    \mathcal{H}(\epsilon\,; \hat{\mathbf{x}}) = \frac{1}{\pi} \mathcal{P} \int_{-\infty}^{\infty} \mathrm{d}\epsilon' 
\frac{\Delta(\epsilon'\,; \hat{\mathbf{x}})}{\epsilon - \epsilon'}
= |V(\hat{\mathbf{x}})|^2 \mathcal{P} \int_{-\infty}^{\infty} \mathrm{d} \epsilon'
\frac{\rho(\epsilon')}{\epsilon - \epsilon'}.
\end{equation}

Because MDEF treats nuclear motion classically, we drop the operator hat on the configurational vector $\hat{\textbf{x}} \rightarrow \textbf{x}$ for the remainder of the paper. Also, the projected electron–mode coupling reduces in the single-resonance limit to the energy-resolved scalar
\begin{equation}
\begin{split}
    \mathcal{M}^{\mu}(\epsilon\,;\mathbf{x}) = {} & \left( \frac{\partial h(\mathbf{x}) + \mathcal{H}(\epsilon\,;\mathbf{x})}{\partial x_\mu} \right. \\
    & + \left. \frac{\epsilon-h(\mathbf{x})-\mathcal{H}(\epsilon\,;\mathbf{x})}{\Delta(\epsilon\,;\mathbf{x})}\frac{\partial \Delta(\epsilon\,;\mathbf{x})}{\partial x_\mu} \right), 
\end{split}
\end{equation}
when following the projection of the orbital trace onto the impurity spectral function as in Brandbyge \textit{et al.}. in \cite{brandbyge_1995_electronically}; the explicit reduction is given in Appendix \,\ref{ap:multi-orbital-to-single-energy-level}. 
The frequency-dependent tensorial friction of NAH for multiple degrees of freedom is given by
\begin{equation}
\label{eq:Anderson-memory-friction}
\begin{aligned}
    \mathcal{K}_{\mu\nu}(\omega\,;\mathbf{x}) &= \Lambda_{\mu\nu}(\omega\,;\mathbf{x}) /2 \\ &= -\frac{1}{4\omega} \left[\operatorname{Im}\Pi^r_{\mu\nu}(\omega\,;\mathbf{x}) + \operatorname{Im}\Pi^r_{\nu\mu}(\omega\,;\mathbf{x})\right],
\end{aligned}
\end{equation}
where $\operatorname{Im}\Pi_{\mu\nu}^{r}$  is
\begin{equation}
\begin{split}
    \operatorname{Im}\Pi^r_{\mu\nu}(\omega\,;\mathbf{x}) 
    &= \int^{\infty}_{-\infty}\frac{\mathrm{d}\epsilon}{2\pi}\; A^\mu(\epsilon\,;\mathbf{x}) \\
    &\quad \times A^\nu(\hbar\omega+\epsilon\,;\mathbf{x})\, \bigl[n_F(\hbar\omega+\epsilon) - n_F(\epsilon)\bigr].
\end{split}
\end{equation}
The vertex functions $A^\mu(\epsilon\,;\mathbf{x})$ take the general form
\begin{equation}
\label{eq:Ak_old}
\begin{split}
A^\mu(\epsilon\,;\mathbf{x}) &= \mathcal{A}(\epsilon\,;\mathbf{x}) \mathcal{M}^{\mu}(\epsilon\,;\mathbf{x}) \\
&= \mathcal{A}(\epsilon\,;\mathbf{x}) \biggl[ \frac{\partial (h(\mathbf{x}) + \mathcal{H}(\epsilon\,;\mathbf{x}))}{\partial x_\mu} \\
&\qquad + \frac{\epsilon - h(\mathbf{x}) - \mathcal{H}(\epsilon\,;\mathbf{x})}{\Delta(\epsilon\,;\mathbf{x})} \frac{\partial \Delta(\epsilon\,;\mathbf{x})}{\partial x_\mu} \biggr].
\end{split}
\end{equation}
Under the product ansatz introduced in \cref{eq:constant-hybridistaion}, the hybridization and energy shift satisfy the logarithmic derivative relations
\begin{equation}
    \frac{\partial \Delta}{\partial \mathbf{x}} = \frac{\partial \ln |V(\mathbf{x})|^2}{\partial \mathbf{x}} \Delta, \quad 
    \frac{\partial \mathcal{H}}{\partial \mathbf{x}} = \frac{\partial \ln |V(\mathbf{x})|^2}{\partial \mathbf{x}} \mathcal{H}.
\end{equation}
Substituting these into \cref{eq:Ak_old}, the terms involving the energy shift $\mathcal{H}$ cancel, simplifying the vertex function to
\begin{equation}
    \label{eq:Ak_new}
    A^\mu(\epsilon\,;\mathbf{x}) = \mathcal{A}(\epsilon\,;\mathbf{x}) \left[ \frac{\partial h(\mathbf{x})}{\partial x_\mu} + \frac{\epsilon - h(\mathbf{x})}{\Delta(\epsilon\,;\mathbf{x})} \frac{\partial \Delta(\epsilon\,;\mathbf{x})}{\partial x_\mu} \right].
\end{equation}
By construction, \cref{eq:Anderson-memory-friction} ensures that the friction matrix remains symmetric under the exchange of indices $\mu$ and $\nu$.

A widely used approximation is the wide-band limit (WBL), in which the reservoir density of states is assumed to be energy independent,
\[
\rho(\epsilon) \rightarrow c.
\]
Consequently, the hybridization function also becomes energy independent,
\[
\Delta(\epsilon\,;\mathbf{x})
\rightarrow
\Delta(\mathbf{x}),
\]
and the impurity spectral function takes the Lorentzian form
\begin{equation}
\label{eq:Ja-Lorentzian}
\mathcal{A}_{a}(\epsilon\,;\mathbf{x})
=
\frac{2\Delta(\mathbf{x})}
{\left[\epsilon-h(\mathbf{x})\right]^2
+\Delta(\mathbf{x})^2},
\end{equation}
centered at the impurity energy level, $h(\mathbf{x})$. In the zero-frequency limit, the frequency-dependent friction tensor defined in \cref{eq:Anderson-memory-friction} reduces to
\begin{equation}
\eta(\mathbf{x}) := \lim_{\omega \rightarrow 0} \mathcal{K}(\omega\,;\mathbf{x}),
\end{equation}
thereby recovering the Markovian electronic friction derived by Brandbyge \textit{et al.}\ \cite{brandbyge_1995_electronically} and Jin and Subotnik \cite{jin_practical_2019}.

\subsection{Positive semi-definiteness conditions for frequency-dependent electronic friction}\label{sec:PSDfrequencyfriction}
Recall that the fluctuation-dissipation theorem in MDEF ensures that, at thermal equilibrium, system-bath total energy is conserved. Given the nuclear configuration $\mathbf{x}$ at time $t$ and $s$, \cref{eq:fdt} simplifies to
\begin{equation}
    \langle \xi^c_\mu(t)\,\xi^c_\nu(s) \rangle = k_{\mathrm{B}}T\,\mathcal{K}_{\mu\nu}(\mathbf{x}\,;t-s),
    \label{eq:FDT}
\end{equation}
Total energy conservation requires the correlation kernel to be positive semidefinite (PSD).
\begin{equation}
\sum_{\mu\nu}\int_{-\infty}^{\infty}\int_{-\infty}^{\infty}
        f_\mu(t)\,\langle \xi^c_\mu(t)\,\xi^c_\nu(s) \rangle\,f_\nu(s)\;\mathrm{d}t\,\mathrm{d}s \;\geq\; 0
\end{equation}
for every square-integrable function $f_\mu$. Upon Fourier transformation, this condition is equivalent to requiring the friction spectrum matrix to be PSD,
\begin{equation}
    \mathcal{K}(\omega\,;\mathbf{x}) \succeq 0, \quad \forall\omega,
\end{equation}
which guarantees that the electronic bath contributes no negative dissipation at any frequency.

For the single-coordinate Anderson Hamiltonian, the analytical expressions in \cref{eq:self-energyfriction,eq:Anderson-memory-friction}
reveal that the strict positive semidefiniteness of frequency-dependent friction imposes strong constraints on the coordinate dependence of the impurity parameters with either
\begin{equation}
\label{eq:singlemodePSDcondition}
    \partial_x \Delta(x\,;\epsilon)  = 0 \quad \text{or} \quad |\partial_xh| \gg |\partial_x \Delta| .
\end{equation}
More generally, the friction spectrum exhibits a characteristic threshold frequency:
\begin{equation}
\label{eq:threshold_frequency}
\begin{aligned}
\omega_{*}^{(T=0)}
    &:= \frac{\sqrt{6}}{\hbar}
        \left|
            \frac{\partial_x h}{\partial_x \Delta}\,\Delta - h
        \right|, \\
\omega_{*}
    &:= \sqrt{
            \bigl(\omega_{*}^{(T=0)}\bigr)^2
            + \frac{2\pi^2 (k_\mathrm{B}T)^2}{\hbar^2}
        },
\end{aligned}
\end{equation}
such that the scalar frequency-dependent friction
\begin{equation}
\mathcal{K}(\omega\,; x) \ge 0,
\qquad
|\omega| \le \omega_*.
\end{equation}
The threshold $\omega_*$ consists of the intrinsic coupling information of the impurity-bath and a thermal additive correction. Outside this frequency window, the friction spectrum indicates a breakdown of strict positive semidefiniteness. Nevertheless, a quasi-positive semidefinite memory kernel remains valid within the dynamically relevant range $|\omega| < \omega_*$ at a physical temperature $T$. This condition is likewise satisfied if the hybridization is approximately configuration-independent ($\partial_x\Delta \rightarrow 0$) or if the impurity level’s coordinate dependence is sufficiently strong relative to the hybridization width ($|\partial_x h|\gg |\partial_x\Delta|$).

For scenarios involving multiple degrees of freedom, the general conditions required to ensure the strict PSD property of the memory electronic friction matrix generalize to an extra condition:
\begin{equation}\label{eq:collinear}
    \nabla_{\mathbf{x}}h(\mathbf{x})\, || \, \nabla_{\mathbf{x}}\Delta(\epsilon\,;\mathbf{x}).
\end{equation}
which necessitates that the spatial gradients of the impurity energy level and the hybridization width be collinear. This alignment ensures that the coordinate dependence of both quantities is restricted to the same direction. 
Equivalently, when the symmetric frequency-dependent friction tensor
$\mathcal{K}(\omega;\mathbf{x})$ becomes rank-1, its dissipation acts
through a single channel: the friction collapses onto one dissipative
eigenmode, while the orthogonal subspace is
dissipationless. Consequently, the multi-mode PSD property reduces to the
single-mode condition~\cref{eq:singlemodePSDcondition} applied to the
eigenmode $x_{\text{eigen}}$, with the directional derivative
\begin{equation}
    \frac{\partial}{\partial x_{\text{eigen}}}
    = \mathbf{v}_{\text{eigen}}(\omega;\mathbf{x})\cdot\nabla_{\mathbf{x}},
    \qquad |\mathbf{v}_{\text{eigen}}|=1,
\end{equation}
where $\mathbf{v}_{\text{eigen}}(\omega;\mathbf{x})$ is the unit
eigenvector associated with the single non-zero eigenvalue of
$\mathcal{K}(\omega;\mathbf{x})$. The derivations of
\cref{eq:threshold_frequency,eq:collinear} are given in
Appendix~\ref{app:PSD}.

\subsection{Memory friction kernel from Kohn-Sham Density Functional Theory}

We now discuss how the electronic friction kernel is calculated from the Kohn-Sham states in DFT and how this is implemented in the FHI-aims software package~\cite{maurer16, box_ab_2023}. 
The Newns--Anderson model provides a transparent model Hamiltonian for
discussing memory-dependent electronic friction. In an ab initio treatment,
the model electronic Hamiltonian $\hat{H}_{\mathrm{el}}(\mathbf{x})$ is
replaced by the Kohn--Sham electronic Hamiltonian. 

At a given nuclear configuration, $\mathbf{x}$, the one-particle
Kohn--Sham Hamiltonian is
\begin{equation}
\hat{h}_{\mathrm{KS}}[\rho_{\mathbf{x}};\mathbf{x}]
=
\hat{t}_s
+
\hat{v}_{\mathrm{ext}}(\mathbf{x})
+
\hat{v}_{\mathrm{H}}[\rho_{\mathbf{x}}]
+
\hat{v}_{\mathrm{xc}}[\rho_{\mathbf{x}}],
\end{equation}
where $\hat{t}_s$ is the single-particle kinetic-energy operator,
$\hat{v}_{\mathrm{ext}}$ is the electron--nuclear potential,
$\hat{v}_{\mathrm{H}}$ is the Hartree potential, and
$\hat{v}_{\mathrm{xc}}$ is the exchange--correlation potential and $\rho_\mathbf{x}$ is the electronic-density at configuration $\mathbf{x}$. The
Kohn--Sham states, $\ket{n\kk;\mathbf{x}}$, satisfy
\begin{equation}
\hat{h}_{\mathrm{KS}}[\rho_{\mathbf{x}};\mathbf{x}]
\ket{n\kk;\mathbf{x}}
=
\epsilon_{n\kk}(\mathbf{x})
\ket{n\kk;\mathbf{x}},
\end{equation}
 where $\epsilon$ are the Kohn-Sham eigenvalues, $n$ labels the Kohn--Sham band and $\kk$ the crystal momentum. The corresponding equilibrium occupations are
$n_{\mathrm{F}}(\epsilon_{n\mathbf{k}})$.

In an atom-centred basis, as used in FHI-aims, the Kohn--Sham states are expanded as
\begin{equation}
\ket{n\kk;\mathbf{x}}
=
\sum_i
C^i_{n\kk}(\mathbf{x})
\ket{\chi_{i\kk}(\mathbf{x})}.
\end{equation}

The Kohn--Sham problem then becomes a generalized
eigenvalue problem: 
\begin{equation}
\sum_j
H^{}_{ij}(\kk;\mathbf{x})
C^j_{n\kk}(\mathbf{x})
=
\epsilon_{n\kk}(\mathbf{x})
\sum_j
S_{ij}(\kk;\mathbf{x})
C^j_{n\kk}(\mathbf{x}).
\label{eq:KS-generalized-eigenproblem-spin}
\end{equation}
Here, $\mathbf{H}$ and $\mathbf{S}$ are the Kohn--Sham Hamiltonian and
overlap matrices.  
From here on, we leave out the position-dependence of the eigenvalues for brevity, i.e $\epsilon(\mathbf{x}) \to \epsilon$. 

The Kohn-Sham static electron–nuclear coupling vertex matrix element for a Cartesian displacement of atom $a$ along direction $\kappa$ is
\begin{equation}
\tilde{g}^{a\kappa}_{mn}(\kk;\mathbf{x})
=
\Braket{
m\kk;\mathbf{x}
|
\frac{\partial \hat{h}_{\mathrm{KS}}^{}}{\partial \mathbf{x}_{a\kappa}}
|
n\kk;\mathbf{x}
}.
\label{eq:KS-cartesian-EPC-q0-spin}
\end{equation}
Here $\hat{h}_{\mathrm{KS}}$ is the Kohn--Sham Hamiltonian. 
This is a static, screened derivative of the instantaneous Kohn--Sham
Hamiltonian. Here, ``static'' means that the coupling vertex has no
explicit dependence on the external frequency $\omega$

In a nonorthogonal atom-centred basis, this matrix
element is evaluated as
\begin{equation}
\begin{split}
\tilde{g}_{mn}^{a\kappa,}(\kk;\mathbf{x})
=
\sum_{ij}
\left(C_{m\kk}^{j}\right)^*
C_{n\kk}^{i}
\bigg[
&
H_{ij}^{a\kappa,(1),}(\kk)
\\ - 
\epsilon_{n\kk}
S_{ij}^{a\kappa,L}(\kk)
&
-
\epsilon_{m\kk}
S_{ij}^{a\kappa,R}(\kk)
\bigg].
\end{split}
\label{eq:KS-cartesian-EPC-nao-q0-spin}
\end{equation}
In this expression $\mathbf{H}^{a\kappa,(1),}$ is the 
first-order screened Hamiltonian response matrix, while
$\mathbf{S}^{a\kappa,L/R}$ are the left and right overlap-response matrices arising from the displacement of the atom-centred basis functions.  These quantities can be evaluated with finite difference evaluation or density functional perturbation theory~\cite{maurer16, shang_lattice_2017, box_ab_2023, abbott_roadmap_2026}.
For computational efficiency, we apply the "Head-Gordon-Tully" approximation~\cite{maurer16} and evaluate the matrix element as:
\begin{equation}
\begin{split}
\tilde{g}_{mn}^{a\kappa,}(\kk;\mathbf{x})
\approx
\sum_{ij}
\left(C_{m\kk}^{j}\right)^*
C_{n\kk}^{i}
\bigg[
&
H_{ij}^{a\kappa,(1),}(\kk)
- 
\epsilon_{\mathrm{F}}
S_{ij}^{a\kappa,(1)}(\kk)\bigg]
\end{split}
\label{eq:KS-cartesian-EPC-nao-q0-hgt}
\end{equation}
where $\mathbf{S}^{\alpha,\kappa,(1)}$ is the first-order overlap matrix, and $\epsilon_\mathrm{F}$ is the Fermi level. This approximation has previously been demonstrated to have only minor effects on the coupling calculated~\cite{maurer16}. 

From this point we use the indices $\mu$ and $\nu$ for the nuclear-coordinate
basis in which the friction is analysed. For the NO/Au(111) results
presented later, the relevant coordinates are
$\mu,\nu\in\{r,\theta,z\}$, corresponding to the internal N--O stretch,
the molecular orientation angle relative to the surface normal, and the
centre-of-mass coordinate normal to the surface, respectively. The electron--phonon matrix elements are first evaluated in Cartesian
atomic coordinates; the resulting friction tensors and spectrum are then transformed to the internal-coordinate basis.

Within the local linear-coupling approximation introduced in
\cref{sec:linearisationEHamiltonian}, the dissipative response of the
auxiliary Kohn--Sham system defines a frequency-dependent friction
tensor~\cite{box_ab_2023}
\begin{equation}
\begin{split}
\mathcal{K}_{\mu\nu}^{\mathrm{KS}}
(\omega;\mathbf{x})
={}&
2\pi\hbar
\sum_{mn}
\int_{\mathrm{BZ}}
\frac{\mathrm{d}\mathbf{k}}{\Omega_{\mathrm{BZ}}}\,
\tilde{g}_{mn}^{\mu}
(\mathbf{k};\mathbf{x})
\left[
\tilde{g}_{mn}^{\nu}
(\mathbf{k};\mathbf{x})
\right]^{*}
\\
&\times
\left[
n_{\mathrm{F}}(\epsilon_{n\mathbf{k}})
-
n_{\mathrm{F}}(\epsilon_{m\mathbf{k}})
\right]
\frac{
\delta\!\left(
\epsilon_{m\mathbf{k}}
-
\epsilon_{n\mathbf{k}}
-
\hbar\omega
\right)
}{
\hbar\omega
},
\end{split}
\label{eq:KS-local-frequency-friction-q0}
\end{equation}
The prefactor of two accounts for spin degeneracy in the
spin-unpolarized calculations used here. Because the nuclear
perturbations have wave vector $\mathbf{q}=0$, the electronic
transitions conserve crystal momentum and are vertical in
$\mathbf{k}$ space.
The difference between the Fermi occupation factors describes the net dissipative response of the electronic subsystem
and ensures detailed balance at finite electronic temperature~\cite{maurer16}.

For an exact delta distribution,  the replacement
$\hbar\omega\rightarrow
\epsilon_{m\kk}-\epsilon_{n\kk}$ in the denominator of \cref{eq:KS-local-frequency-friction-q0} 
can  be made~\cite{maurer16,box_ab_2023} which consequently becomes
\begin{equation} 
\begin{split}
&\mathcal{K}_{\mu\nu}^{\mathrm{KS}}
(\omega\,;\mathbf{x})
\approx
2\pi\hbar
\sum_{ mn}
\int
\frac{\mathrm{d}\kk}{\Omega_{\mathrm{BZ}}}
\,
\tilde{g}^{\mu}_{mn}(\kk;\mathbf{x})
\left[
\tilde{g}^{\nu}_{mn}(\kk;\mathbf{x})
\right]^*
\\
&\times
\left(
n_F(\epsilon_{n\kk})
-
n_F(\epsilon_{m\kk})
\right)
\frac{
\delta\!\left(
\epsilon_{m\kk}
-
\epsilon_{n\kk}
-
\hbar\omega
\right)
}{
\epsilon_{m\kk}-\epsilon_{n\kk}
}.
\end{split}
\label{eq:abinitio_lambda}
\end{equation}
In a numerical Brillouin-zone integration, the delta distribution is replaced by a finite-width broadening function (here a Gaussian with width $\sigma$) and \cref{eq:KS-local-frequency-friction-q0,eq:abinitio_lambda} become distinct and must be converged with respect to the
broadening width and $\kk$-point sampling
\cite{box_ab_2023}.

In practice, FHI-aims outputs 
\cref{eq:abinitio_lambda} for $\omega>0$ only. Thus, to connect the frequency-resolved tensor to the causal memory kernel
introduced in \cref{sec:Single position based friction kernel}, we
identify $\mathcal{K}^{\mathrm{KS}}(\omega;\mathbf{x})$ with one half
of the even, two-sided dissipation spectrum:
\begin{equation}
\Lambda_{\mu\nu}^{\mathrm{KS}}
(\omega;\mathbf{x})
=
2\mathcal{K}_{\mu\nu}^{\mathrm{KS}}
(\omega;\mathbf{x}).
\end{equation}
Using \Cref{eq:etat} 
and the evenness of
$\Lambda^{\mathrm{KS}}(\omega;\mathbf{x})$, the causal time-domain
kernel is
\begin{equation}
\mathcal{K}_{\mu\nu}^{\mathrm{KS}}
(t-t';\mathbf{x})
=
\frac{2\Theta(t-t')}{\pi}
\int_{0}^{\infty}
\mathrm{d}\omega\,
\mathcal{K}_{\mu\nu}^{\mathrm{KS}}
(\omega;\mathbf{x})
\cos(\omega (t-t')).
\label{eq:local-K-from-Lambda-cos}
\end{equation}
The factor of two in this cosine transform comes from reconstructing
the even, two-sided spectrum from its positive-frequency branch. 

In practice, the electronic spectrum is evaluated only up to a finite
cutoff $\omega_{\max}=E_{\max}/\hbar$. A hard cutoff would introduce
artificial oscillations in the time-domain kernel, so we multiply the
spectrum by a smooth tapering window $w(\omega)$. The window is unity
over the well-resolved part of the spectrum and decreases continuously
to zero as $\omega\rightarrow\omega_{\max}$. The local kernel used in
the calculations is therefore approximated as
\begin{align}
\mathcal{K}_{\mu\nu}^{\mathrm{KS}}
(t-t';\mathbf{x})
\simeq 
\frac{2\Theta(t-t')}{\pi}
\int_{0}^{\omega_{\max}} 
\mathrm{d}\omega\, & \\
\times w(\omega)\, 
\mathcal{K}_{\mu\nu}^{\mathrm{KS}}
(\omega;\mathbf{x})
\cos(\omega (t-t')).
\label{eq:local-K-from-windowed-spectrum}
\end{align}
We use the same window and cutoff at every configuration,

We can compare to a Markovian reference by taking the zero-frequency limit, $\hbar\omega \to 0$, of \cref{eq:abinitio_lambda}, and we arrive at the commonly employed time-dependent perturbation theory/orbital-dependent friction (ODF)~\cite{hellsing1984,maurer16,spiering2018,box_ab_2023}.
\begin{equation}
\begin{split}
\eta^{\mathrm{ODF}}_{a\kappa,a'\kappa'}(\mathbf{x}) 
&= 
\lim _{\omega \rightarrow 0^{+}} \mathcal{K}^{\mathrm{KS}}(\omega) \\
\end{split}
\end{equation}

In this work, we define another type of  Markovian friction coefficient based on the average of the frequency-dependent friction kernel (\cref{eq:abinitio_lambda}) over a given frequency range $[\omega_1,\omega_2]$:
\begin{equation}
\eta^{\mathrm{Avg}}_{a\kappa,a'\kappa'}(\mathbf{x})
=
\frac{1}{\omega_2-\omega_1}
\int_{\omega_1}^{\omega_2}
\mathcal{K}_{a\kappa,a'\kappa'}^{\mathrm{KS}}
(\omega;\mathbf{x})\,
\mathrm{d}\omega .
\label{eq:eta_avg}
\end{equation}
We weight this friction coefficient towards higher frequencies (to contrast to the lower frequency limit employed in ODF ). This is a heuristic choice of Markovian friction tensor with a magnitude representative of the friction spectrum at finite frequencies to aid our later analysis of the effect of including memory in hyperthermal scattering dynamics.

\subsection{\label{sec:CPA} Memory friction energy dissipation via classical-path approximation} 

In the general non-Markovian MDEF dynamics discussed in \cref{sec:MDEF}, the total energy variation in a particular nuclear mode $\mu$ up to a finite time $t$ is obtained by multiplying its velocity $\dot{x}_\mu(t)$ and  integrating over time in \cref{eq:MDEF-kmode}:
\begin{equation}
\label{eq:generalDeltaE}
\begin{split}
    \Delta E_{\mu}(t) = {} & -\int^{t}_{0}\mathrm{d}\tau\,\dot{x}_{\mu}(\tau)\sum_{\nu}\int^{\tau}_{0}\mathrm{d}\tau' \\
    &\times \mathcal{K}_{\mu\nu}\bigl( \tau-\tau'\,;\mathbf{x}(\tau), \mathbf{x}(\tau') \bigr)\dot{x}_{\nu}(\tau') \\
    & + \int^{t}_{0}\mathrm{d}\tau\, \dot{x}_{\mu}(\tau) \xi_\mu^{\mathrm{c}}(\tau)
\end{split}
\end{equation}
The first double integral term on the right-hand side describes the dissipative damping originating from the memory-friction kernel, while the last term measures the energy fed back from the electronic bath into the system.

The classical-path approximation (CPA) estimates this memory-friction effect without explicitly propagating the full memory-dependent MDEF dynamics. The first step is to simulate a Born--Oppenheimer molecular dynamics (BOMD) trajectory $\{\mathbf{x}^{\text{BO}},\dot{\mathbf{x}}^{\text{BO}}\}$ governed by the classical equation of motion:
\begin{equation}
\mathbf{M} \ddot{\mathbf{x}}^{\mathrm{BO}} = -\nabla_{\mathbf{x}^{\mathrm{BO}}} E_0.
\end{equation}
The frictional energy dissipation along the trajectory in mode $\mu$ is then estimated as
\begin{equation}
\begin{split}
\Delta E^{f}_\mu(t) := \int^{t}_{0}\mathrm{d}\tau\,\dot{x}^{\mathrm{BO}}_{\mu}(\tau)\sum_{\nu}\int^{\tau}_{0}\mathrm{d}\tau' \\
\times\,\mathcal{K}^{\text{BO}}_{\mu\nu}\bigl( \tau,\tau' \bigr)\,\dot{x}^{\mathrm{BO}}_{\nu}(\tau')
\end{split}
\end{equation}
This evaluates the dissipative work on an unchanged Born–Oppenheimer path and therefore neglects friction-induced steering, changes in residence time, and stochastic energy transfer.

In the electronic Hamiltonian linearization regime discussed in \cref{sec:linearisationEHamiltonian}, we obtain the \textit{local CPA memory kernel}, 
\begin{equation}
\label{eq:local-phase-point}
\mathcal{K}^{\text{BO}}_{\mu\nu}\bigl( \tau,\tau' \bigr) := \mathcal{K}_{\mu\nu}\bigl( \tau-\tau'\,;\mathbf{x}^{\mathrm{BO}}(\tau) \bigr), 
\end{equation} for both the Anderson-type and full dimensional Kohn-Sham calculations.

The kernel \cref{eq:local-phase-point} is evaluated at one configuration along the
trajectory, but the nonlocal kernel it approximates depends on both endpoints
$\mathbf{x}^{\mathrm{BO}}(\tau)$ and $\mathbf{x}^{\mathrm{BO}}(\tau')$. Collapsing to a single
configuration therefore requires a choice of reference point, and since generically
$\mathbf{x}^{\mathrm{BO}}(\tau)\neq\mathbf{x}^{\mathrm{BO}}(\tau')$, the following inequality holds: 
\begin{equation}
\mathcal{K}_{\mu\nu}\bigl(\tau-\tau'\,;\mathbf{x}^{\mathrm{BO}}(\tau)\bigr)
\neq
\mathcal{K}_{\mu\nu}\bigl(\tau-\tau'\,;\mathbf{x}^{\mathrm{BO}}(\tau')\bigr),
\end{equation}
leaving the estimator dependent on an arbitrary anchoring of the kernel to one end of the
trajectory segment. To assess the effect of this choice and with the aim of removing this ambiguity, we define the \textit{arithmetic mean CPA memory kernel}:
\begin{equation}
\label{eq:arithmeticMean}
\mathcal{K}^{\mathrm{BO}}_{\mu\nu}(\tau,\tau')
= \tfrac{1}{2}\Bigl[
\mathcal{K}_{\mu\nu}\bigl(\tau-\tau'\,;\mathbf{x}^{\mathrm{BO}}(\tau)\bigr)
+\mathcal{K}_{\mu\nu}\bigl(\tau-\tau'\,;\mathbf{x}^{\mathrm{BO}}(\tau')\bigr)
\Bigr],
\end{equation}
which is the component of the local kernel that is invariant under exchange of the two
endpoint configurations $\mathbf{x}^{\mathrm{BO}}(\tau)\leftrightarrow\mathbf{x}^{\mathrm{BO}}(\tau')$.
This endpoint-exchange symmetrization weights both configurations equally, restoring the
symmetry of the nonlocal kernel in its configurational arguments while leaving the temporal
memory $\tau-\tau'$ and the mode-index structure unchanged. The main results of this work use the local CPA memory kernel, while comparison between both defined memory kernels are presented in the SM, showing very little sensitivity of the results with respect to this choice. For all considered trajectories, the projectile molecule begins sufficiently far from the surface such that the electronic friction is negligible at this point. Therefore, truncating the history at the start of the trajectory does not omit any appreciable dissipative energy.

\section{Computational Methods}

\begin{figure}
    \centering
    \includegraphics[width=3.3in]{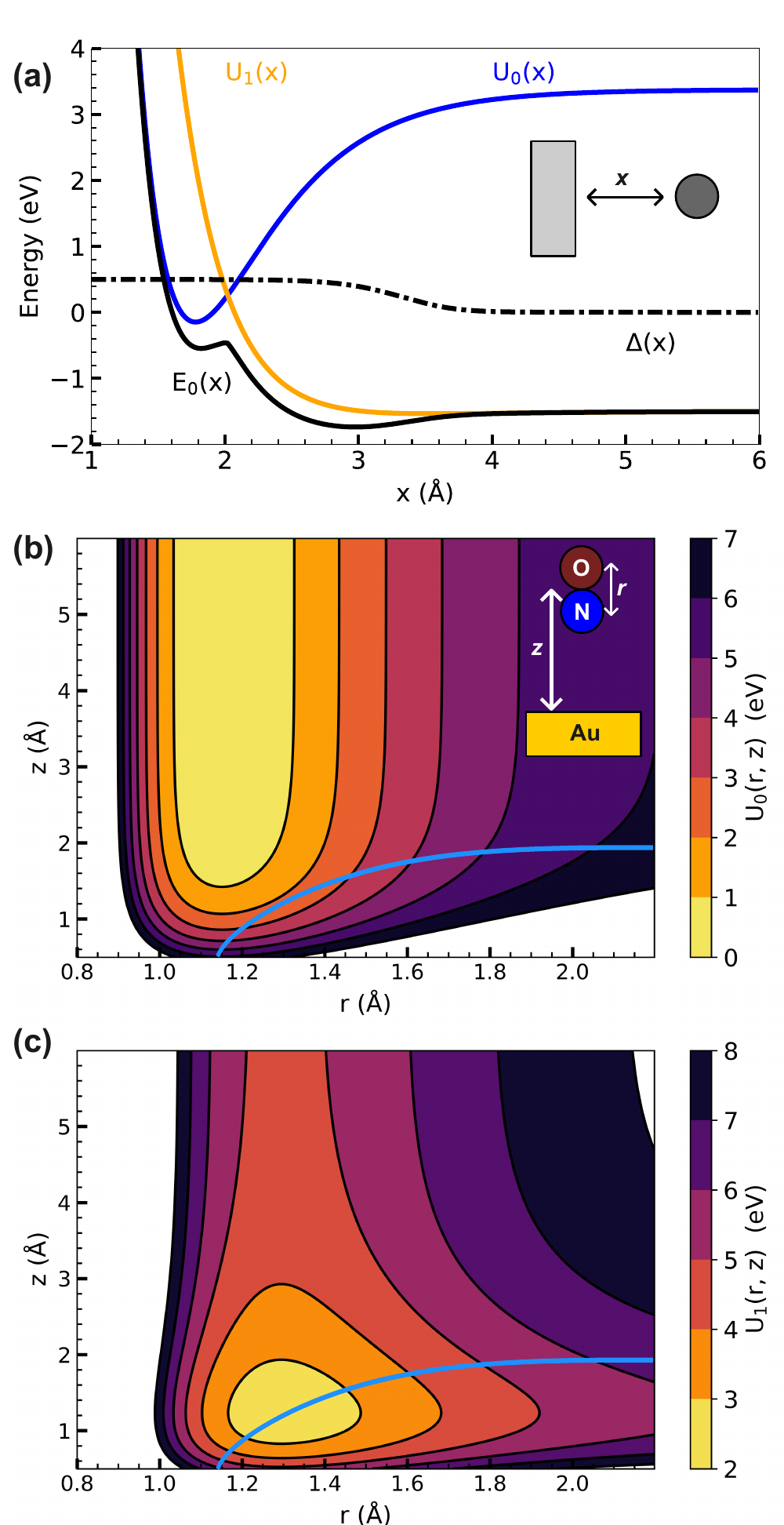}
    \caption{
    Panel (a): Energy curves of the one-dimensional model introduced by Erpenbeck and Thoss~\cite{erpenbeck2018}. $U_0(x)$ and $U_1(x)$ represent the diabatic energy curves of the neutral and charged particle interacting with a metal surface. The black dashed line represents the hybridization function $\Delta(x)$, whereas the black solid curve stands for the adiabatic ground state energy, $E_0(x)$, of the NAH constructed with the three aforementioned quantities. Panel (b): Two-dimensional diabatic energy landscape of a neutral NO molecule at a Au(111) surface, $U_0(r, z)$, as a function of the bond length, $r$, and the height of the centre-of-mass of the molecule from the surface~\citenum{gardner2023a}. Panel (c): Diabatic energy landscape of the NO anion at a Au(111) surface. The blue line in panel (b) and (c) indicates the contour where the two diabatic landscapes intersect.}
    \label{fig:Systems}
\end{figure}

\subsection{Erpenbeck--Thoss model Hamiltonian}
Erpenbeck \textit{et al.} \cite{erpenbeck2018} introduced a two-state impurity Anderson-type Hamiltonian to study current transport in a molecule-metal junction, which we will call Erpenbeck-Thoss (ET) model. Gardner \textit{et al.} \cite{gardner2023} previously adopted the model to study nonadiabatic energy transfer during scattering of an adsorbate impurity from a metal surface.

The diabatic potential energy surface of the unoccupied impurity state is defined as 
\begin{equation}
\label{eq:ETU0}
    U_0(x) = D_e \left( e^{-a(x-x_0)} - 1 \right)^2 + c.
\end{equation}
The potential for the charged impurity state becomes repulsive: 
\begin{equation}
\label{eq:ETU1}
    U_1(x) = D_1 e^{-2a'(x-x_0)} - D_2 e^{-a'(x-x_0)} + V_{\infty}.
\end{equation}
Within the wide-band limit, the hybridisation function between molecule and surface is 
\begin{equation}
\label{eq:ETChem}
    \Delta(x) = \Delta_0 \left( \frac{1 - q}{2} \left[ 1 - \tanh\left( \frac{x - \tilde{x}}{\tilde{a}} \right)\right] + q \right)^2,
\end{equation}
where the prefactor $\Delta_0$ is a tunable parameter that defines the strength of the impurity-bath coupling. \Cref{fig:Systems}(a) illustrates the energy landscapes of the ET model. The parameters for \cref{eq:ETU0,eq:ETU1,eq:ETChem} are given in Table SI in the SM.

\subsection{Model Hamiltonian for NO/Au(111)}
Based on data calculated with constraint DFT by Meng and Jiang \cite{meng2022}, Gardner, Habershon and Maurer \cite{gardner2023a} parametrised a reduced two-dimensional NAH description of NO scattering from Au(111) in the wide band limit, which we shall call the Gardner--Habershon--Maurer (GHM) model.  The model assumes that the molecular axis remains perpendicular to the surface with N facing down and includes only two nuclear coordinates: the center-of-mass molecule–surface distance, $z$, and the internal N-O bond length, $r$.

The form of the neutral and charged diabatic potential energy surfaces for the NO molecule are chosen to be 
\begin{equation}
\label{eq:NOAuU0}
    U_0(r, z) = V_{\text{M}}[r - r_0; D_0, a_0] + \bar{D}_0\exp[-b_0(z - z_0)] + c_0
\end{equation}
\begin{equation}
\label{eq:NOAuU1}
        U_1(r, z) = V_{\text{M}}[r - r_1; D_1, a_1] + V_{\text{M}}[z - z_1; D_2, a_2] + c_1
\end{equation}
where $V_{\text{M}}$ is the Morse potential defined as
\begin{equation}
\label{eq:NOAuVm}
    V_{\text{M}}(x; D, a) = D[\exp(-2ax) - 2\exp(-ax)].
\end{equation}
The coupling is chosen to depend only on the molecule-surface distance
\begin{equation}
\label{eq:NOAuChem}
    \Delta(z) = \Delta_0[1-\tanh{(z/\tilde{a})}]^2.
\end{equation}
The full set of parameters for \cref{eq:NOAuU0,eq:NOAuU1,eq:NOAuVm,eq:NOAuChem} is given in Table SII in the SM. In contrast to the ET model, the impurity-metal coupling  $\Delta_0$ is treated as a fixed parameter, defined by the parametristaion against DFT data. The schematic of the GHM model and the corresponding potential energy surfaces $U_1$ and $U_0$ are shown in \cref{fig:Systems}(b--c).

\subsection{DFT calculations and  trajectory analysis for NO/Au(111)}

In addition to the two-dimensional GHM model, we analyze memory electronic friction along high-dimensional molecular dynamics trajectories of highly vibrationally excited \ce{NO} scattering on a  Au(111) surface.  
Classical trajectories were propagated with the VENUS chemical dynamics code~\cite{hu_vectorization_1991} on the full-dimensional PES proposed in  Ref.~\cite{yin_strong_2019}. Details on the quality of the PES can be found in Ref.~\cite{yin_strong_2019}. 
Six trajectories with $v_i{=}16$, $J_i{=}0$ and $E_i=0.52$~eV were generated using Born-Oppenheimer molecular dynamics. A  visualisation of these six trajectories can be found in the SM (Figure S5). 
The surface was modeled as a four-layer $3\times3$ Au(111) slab with the bottom two layers kept fixed.  
Friction kernels, $\boldsymbol{\mathcal{K}}(\omega)$,  were obtained from finite-difference calculations in \textsc{FHI-aims}~\cite{box_ab_2023} using almost identical settings to Ref.~\cite{box_2020_determining} with the exception of using a 2020 light basis instead of a pre-2020 tight basis definition (for information on these definitions, see Ref.~\cite{blum_ab_2009}, which we find to negligibly affect the calculated friction values. Analysis of the numerical stability of these choices for the friction evaluation can be found in Ref.~\cite{box_2020_determining}. The friction spectrum was generated using a Gaussian smearing width of $0.01$~eV, which was additionally broadened to $0.05$~eV for the memory kernel treatment unless stated otherwise.  We analyze the work associated with the electronic friction force along fixed trajectories (stochastic noise omitted) to study the effect of memory in the dissipative kernel. We use the local endpoint treatment as discussed above. Further data on the numerical stability with regards to choices in the trajectory analysis can be found in the SM (Figure S6, Figure S7 and Figure S8).

\section{Results}

\subsection{Electronic memory friction from Newns--Anderson Hamiltonian}

\begin{figure}
    \centering
    \includegraphics[width=1.0\linewidth]{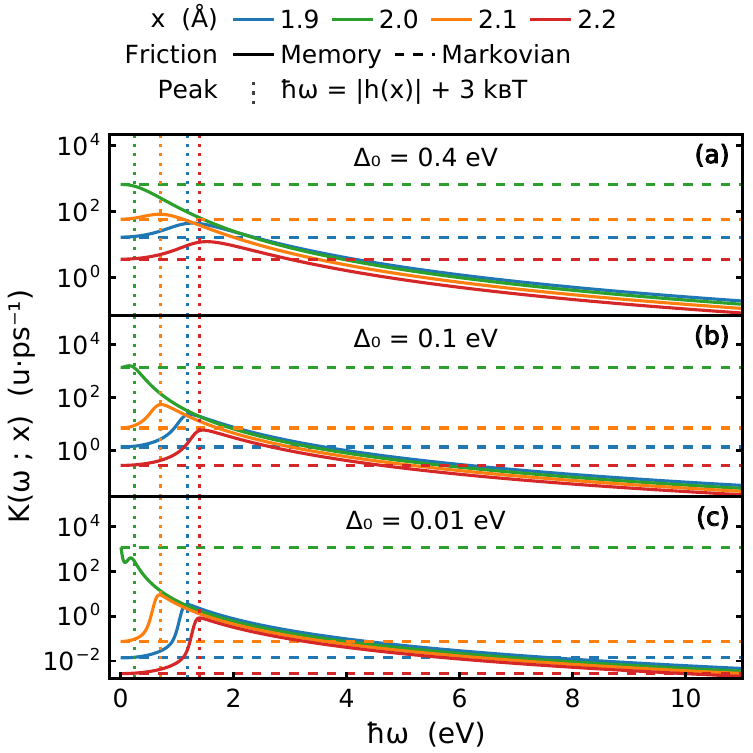}
\caption{%
  Frequency-dependent electronic friction $\mathcal{K}(\omega;x)$ (solid) and its
  Markovian limit (dashed, $\omega$-independent) for the Erpenbeck--Thoss model
  \cite{erpenbeck2018,gardner2023} at $T = 300$~K, versus frequency $\omega$  (axis $\hbar\omega$ in eV; vertical scale logarithmic, friction in  $\mathrm{u\cdot\,ps^{-1}}$). Colors denote nuclear position $x = 1.9$--$2.2$~\AA;  panels (top to bottom) show coupling 
  $\Delta(x{=}0) = 0.4$, $0.1$, $0.01$~eV. Vertical dotted lines mark the particle--hole threshold $|h(x)|+3k_\text{B}T$, where the adsorbate level $h(x)$ crosses the Fermi level and electronic friction induced by population transfer reaches a maximum.
}
    \label{fig:ErpenbeckThossFrequencyfriction}
\end{figure}

The ET model in the wide-band limit~\cite{erpenbeck2018,gardner2023}, describing one-dimensional molecular scattering from a surface, is the first model for which we investigate the electronic friction kernel. \Cref{fig:ErpenbeckThossFrequencyfriction} shows that the frequency-dependent electronic friction spectra $\mathcal{K}(\omega\,;x)$ (solid) exhibit a peaked structure, where, over a wide energy range, the electronic friction first rises above the Markovian, zero-frequency limit (horizontal dashed lines) before decaying to zero. The corresponding kernel in the time-domain corresponds to underdamped system-bath dynamics. We consider three heights of the impurity level above the surface, $x = 1.9, 2.1$ and $2.2$\,\AA , and three values for the impurity-metal coupling, $\Delta_0$.

The strongest nonadiabatic coupling between the impurity and the electronic bath can be found at the configuration where the occupied and unoccupied states $U_1$ and $U_0$ cross ($x\approx2$\,\AA, green curves), as it represents the position at which the impurity energy level crosses the Fermi level of the metal. The corresponding curves accordingly exhibit a monotonic decay and, in this geometry, the corresponding time-domain kernel corresponds to close-to-overdamped dynamics. In this case, the Markovian zero-frequency limit represents the strongest coupling. The peak positions are marked by vertical dotted lines at $|h(x)| + 3k_{\mathrm{B}}T$, which marks the energy separation between the impurity level and an additional contribution due to the finite-temperature Fermi--Dirac population of the states. Excitation of electrons with an energy of $|h(x)| + 3k_{\mathrm{B}}T$ leads to a maximum transfer of population onto the impurity level, as $n_F(-3k_{\mathrm{B}}T) \approx 0.95$ reflects roughly 95\% Fermi electronic population saturation at finite temperature. From the top panel (a) to the bottom (c), the coupling, $\Delta_0$, decreases from 0.4\,eV to 0.01\,eV and the peaks become sharper as the Lorentzian broadening of the impurity due to hybridisation gets smaller and smaller. The same frequency-dependent friction data are represented in contour plots in the SM (Figure S1).

\begin{figure*}[t]
    \centering
    \includegraphics[width=1.0\linewidth]{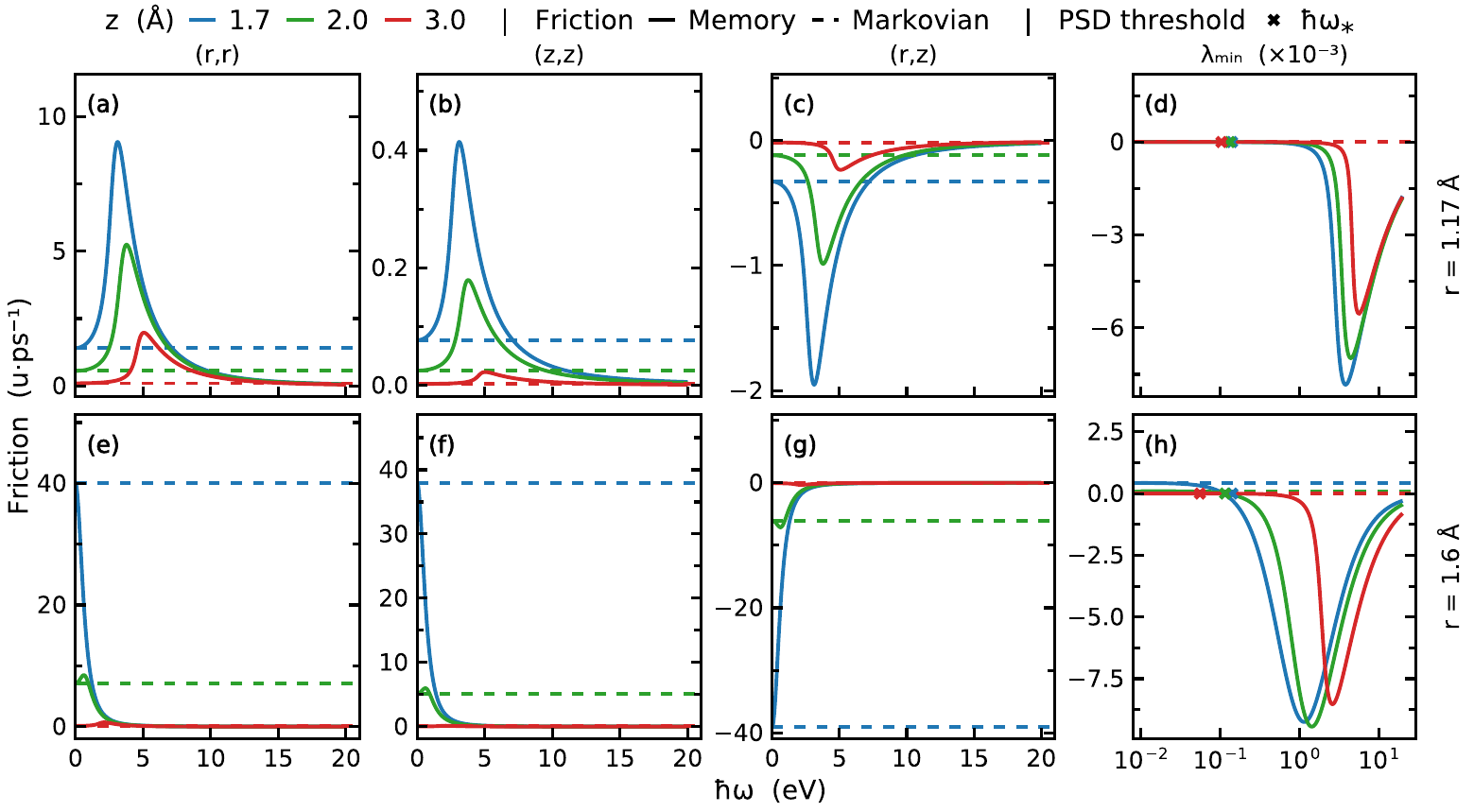}
\caption{
Frequency-dependent electronic friction tensor
$\mathcal{K}_{\mu\nu}(\omega,x)$ (solid) and its Markovian limit (dashed) for NO
scattering from Au(111), described by the Newns--Anderson model Hamiltonian of
Gardner, Habershon, and Maurer~\cite{gardner2023a} in the wide-band limit at
electronic temperature $T = 300$~K. The first three columns are the matrix
elements $\mathcal{K}_{rr}$ (a, e), $\mathcal{K}_{zz}$ (b, f) and the off-diagonal coupling
$\mathcal{K}_{rz}$ (c, g) on a linear $\hbar\omega$ axis; the fourth is the smallest
eigenvalue $\lambda_{\min}(\mathcal{K})$ (note
the $\times 10^{-3}$ scale), shown on a logarithmic $\hbar\omega$ axis.
Crosses ($\times$) in the fourth column mark the analytic threshold frequency
$\hbar\omega_{*}$ given by \cref{eq:threshold_frequency} for each $z$, indicating with the zero crossing of
$\lambda_{\min}$. Rows: two N--O bond lengths $r$ ($1.17$ and $1.7$~\AA); colors: three molecule--surface distances $z$ ($1.7$, $2.0$, $3.0$~\AA).}
    \label{fig:ElementwiseNOAuNAH_friction}
\end{figure*}

The second Anderson-type Hamiltonian we study is the GHM model \cite{gardner2023a}, a reduced two-dimensional description of NO scattering from Au(111) that retains two nuclear coordinates: the molecule–surface distance $z$ and the internal N-O bond length $r$. \Cref{fig:ElementwiseNOAuNAH_friction} shows the frequency-dependent (solid) and Markovian (dashed) electronic friction tensors evaluated at fixed configurations spanning $z=1.7, 2.0, 3.0$\,\AA{} and $r = 1.17, 1.6$\,\AA. These values were chosen to probe friction across the coordinates that are relevant for the scattering process, from the molecule in its equilibrium configuration ($r=1.17$\,\AA{}) far from the surface to the stretched ($r=1.7$\,\AA{}), near-surface geometry where nonadiabatic electron transfer is strongest. As reported by Meng and Jiang \cite{meng2022}, $U_0$ and $U_1$ cross at $r = 1.6$\,\AA{} and $z=1.7$\,\AA{}. As a result, friction is much stronger in the lower panels (e -- h) and the maximum friction is reached in the zero-frequency limit rather than at a finite frequency. 

At the equilibrium bond length $r = 1.17$\,\AA{}, the diagonal elements of the frequency-dependent friction matrix rise from their $\omega\!\to\!0$ (Markovian) values to a pronounced peak near $\hbar\omega \approx 2$--$3$\,eV before decaying to zero [\Cref{fig:ElementwiseNOAuNAH_friction}(a)--(c)]. The vibrational element $\mathcal{K}_{rr}$ dominates: its peak exceeds that of the translational element $\mathcal{K}_{zz}$, indicating that nonadiabatic friction and vibrational energy dissipation are highly anisotropic and dominated by intramolecular vibration. This is well known and related to the fact that vibrational motion more strongly perturbs the electronic structure than molecule-surface motion \cite{Head-GordonTully1992,askerka16,maurer16}. The off-diagonal $(r,z)$ component of the frequency-dependent friction tensor shows a negative-valued peak that reflects the nonadiabatically induced kinematic coupling between the two degrees of freedom of the model. At finite perturbing frequencies of 2 -- 5~eV, which can be induced through high translational incidence energies or high initial vibrational excitation, friction is significantly enhanced at $r=1.17$\,\AA{}. In this case, the Markovian approximation would likely heavily underestimate frictional energy dissipation. 

For the configuration $r = 1.6$\,\AA{}, $z = 1.7$\,\AA{} that corresponds to the diabatic crossing geometry, the frequency-dependent friction is maximal at $\omega = 0$ and decays monotonically $\approx 2.5$\,eV (\cref{fig:ElementwiseNOAuNAH_friction}(e)--(g)). Its overall magnitude is several times larger than friction at other geometries. Both degrees of freedom yield friction that is of roughly equal magnitude; however, the off-diagonal is also very large, and therefore the eigenmodes of the friction tensor will yield highly anisotropic friction. In this geometry, at finite perturbing frequencies, memory-dependent friction would yield lower energy dissipation than Markovian friction. Therefore, the Markov approximation has the potential to over- or underestimate friction depending on the geometry.

In addition to the vibrational and translational components of the friction tensor and their coupling, $\mathcal{K}_{rr}$ and $\mathcal{K}_{zz}$ (panels b,f), $\mathcal{K}_{rz}$, \cref{fig:ElementwiseNOAuNAH_friction}, panels d and h also report the value of the smallest eigenvalue $\lambda_{\text{min}}(\mathcal{K})$, i.e. the eigenmode that is least affected by damping due to electronic friction. A PSD friction tensor has $\lambda_{\min}(\mathcal{K}) \ge 0$ at every frequency, guaranteeing that coupling to the bath only yields dissipation. In the GHM model, the Markovian tensor (dashed) remains PSD at all geometries, whereas a small negative eigenvalue emerges for the frequency-dependent tensor (solid), reaching
$\lambda_{\min} \approx -8\times10^{-3}\,\mathrm{u\,ps^{-1}}$ depending on
$z$ and $r$. While this value is negligible in terms of overall magnitude compared to the other components of the tensor, it means that the model does not obey PSD in its given parametrisation based on first-principles data. The crosses indicated in \cref{fig:ElementwiseNOAuNAH_friction} panels d and h mark the analytic threshold frequency $\omega_*$ (\cref{eq:threshold_frequency}). Beyond $\omega_*$ the friction ceases to act as a purely dissipative
term, suggesting potentially that MDEF with or without memory may not be an appropriate approximation for this system. Previous benchmark results against experiment \cite{box_2020_determining},  explicit mixed quantum-classical dynamics methods \cite{gardner2023a}, and numerically exact quantum dynamics simulations corroborate this view, although it remains unclear if the inclusion of memory can extend the realm of applicability of MDEF \cite{preston2025}.

\subsection{Energy dissipation in the classical path approximation for Anderson-type models}

\begin{figure}
    \centering
    \includegraphics[width=1.0\linewidth]{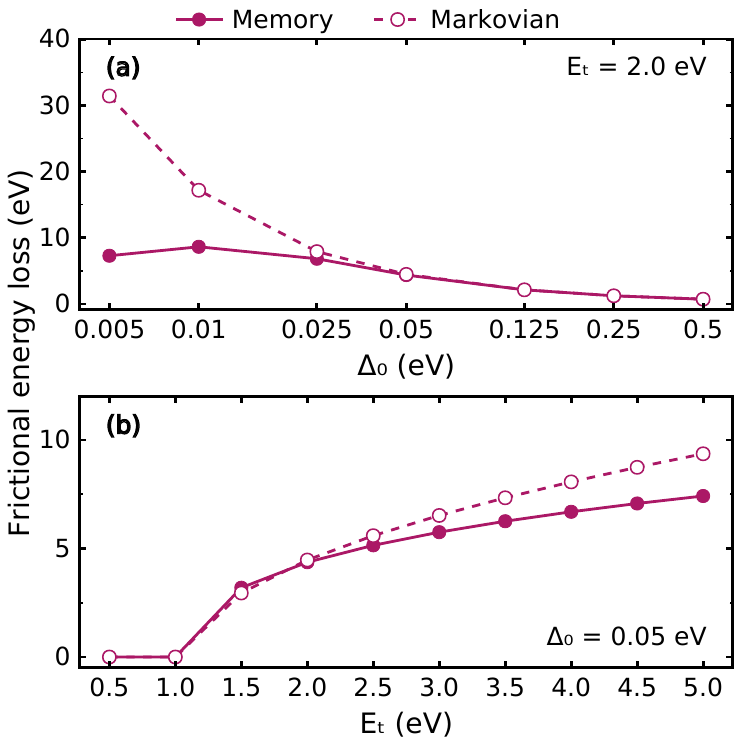}
    \caption{Frictional energy loss, $\Delta E^f$, of impurity scattering, computed within the
    classical-path approximation (CPA), for the Erpenbeck--Thoss
    model~\cite{erpenbeck2018}. (a) $\Delta E^f$ versus the impurity--metal coupling
    strength, $\Delta_0$, at fixed incident translational energy
    $E_{\mathrm{t}} = 2$~eV. (b) $\Delta E^f$ versus incident translational energy,
    $E_{\mathrm{t}}$, at fixed coupling strength $\Delta_0 = 0.05$~eV. In both panels
    the impurity position is initialised at $5$~\AA{} The frictional energy loss for the non-Markovian case with local
    kernel averaging (solid lines, filled circles) is compared with the Markovian
    limit (dashed lines, open circles).
    }
    \label{fig:ErpenbeckThossCPA}
\end{figure}

\paragraph{Erpenbeck--Thoss model.} Non-Markovian effects on the dissipative dynamics can be assessed by calculating the frictional energy loss along a classical molecular dynamics trajectory. The chosen initial kinetic energy of $E_t = 2.0$\,eV is sufficient to reach the maximum of friction at all considered impurity-metal coupling strengths,  $\Delta_0$, ranging from 0.005 to 0.5\,eV (\cref{fig:ErpenbeckThossCPA}).  The Markovian frictional energy loss (open circles) increases with decreasing hybridisation strength. This behavior reflects the fact that dissipation occurs predominantly when the trajectory reaches the position near $2$ \AA{}, where $h=0$ and the Markovian friction reaches a maximum. Specifically, Markovian friction in the ET model, obtained from the zero-frequency limit of \cref{eq:Anderson-memory-friction}, scales as
$$
\eta \propto \frac{\Delta^2}{(h^2+\Delta^2)^2}
$$
At the crossing point $h(x) = 0$, this reduces to 
$$
\eta \propto 1 / \Delta^2,
$$
which gives rise to the well-known stopping power singularity of electronic friction. Even at $\Delta_0= 0.05$\,eV, the total frictional energy loss is sufficient to dissipate all kinetic energy and stop the particle before it can cross the ground-state barrier at 2\,\AA{} \cite{gardner2023}.

The memory-dependent frictional energy dissipation computed along the same trajectory with the local averaging kernel (Eq.\ \eqref{eq:local-phase-point}) only starts to deviate from the Markovian result for coupling values $\Delta_0< 0.05$\,eV. At larger values, the memory-dependent dissipation converges to its Markovian counterpart. This phenomenon arises as the frequency-dependent friction broadens with increasing hybridization of the impurity state and memory effects do not lead to an enhancement of the friction magnitude. In this limit (\cref{fig:ErpenbeckThossFrequencyfriction}, panel a), the friction spectrum is well approximated by the constant  spectrum of the Markovian friction. 

In the ET model with small $\Delta_0$, memory effects lead to a reduction of energy loss as the velocity profile of the trajectory samples high-frequency friction values which yield suboptimal dissipation when compared to the Markovian case. An increase in kinetic energy of the particle significantly lowers the effective friction at $x=2$\,\AA{} and slightly increases effective friction at positions above and below $x=2$\,\AA{}, leading to an overall decrease of frictional energy loss along the CPA trajectory. The memory effect correctly describes the time-lag of the electrons as they cannot respond instantaneously to fast projectiles.

In the SM, we show that the choice between the arithmetic and local-averaging kernels (\cref{eq:arithmeticMean,eq:local-phase-point}) for computing the frictional dissipation within the CPA of the ET model has a negligible effect (Figure S2). 


\begin{figure}
    \centering
\includegraphics[width=1.0\linewidth]{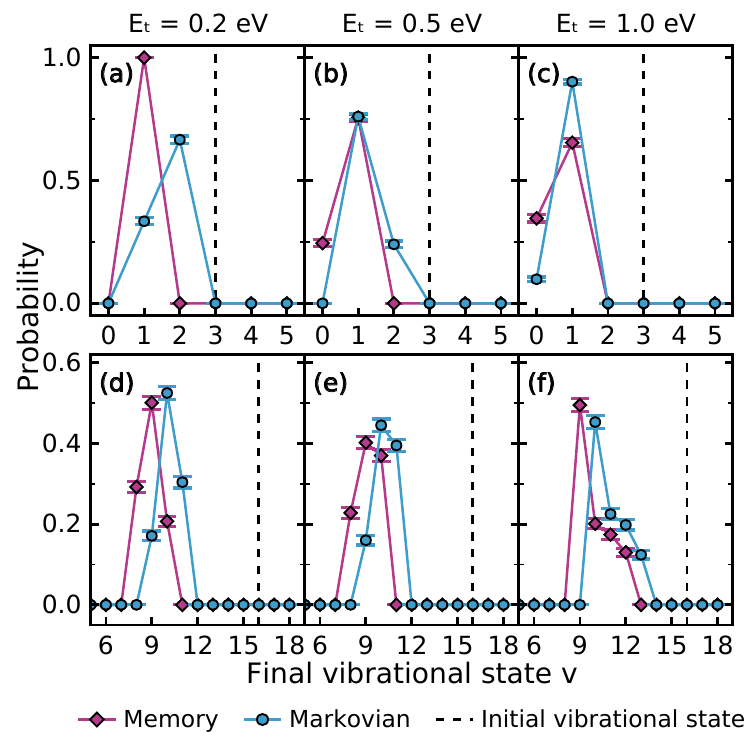}
\caption{Final vibrational-state distributions $P(v_f)$ at $T = 300$~K for
initial states $v_i = 3$ (top row, panels a-c) and $v_i = 16$ (bottom row, panels d-f) and incident
translational energies $E_t = 0.2$, $0.5$, and $1.0$~eV (columns). Markovian and memory friction with the local kernel averaging \cref{eq:local-phase-point} are compared. Each distribution $P(v_f)$ is averaged over 1000 trajectories; error bars indicate the binomial standard error. Dashed vertical lines mark $v_i$.}
    \label{fig:GardnerMaurerCPA}
\end{figure}

\paragraph{Gardner--Habershon--Maurer model for NO/Au(111).}

Based on the results in \cref{fig:ElementwiseNOAuNAH_friction}, we can expect that the vibrational coupling channel dominates electronic friction in the GHM model. Indeed, as shown in Figure S3 where vibrational and translational energy losses calculated using CPA are compared, energy dissipation along $z$ is much weaker. Following previous work \cite{gardner2023a}, we study the vibrational final state distribution upon scattering. We consider an initial vibrational state $v_i=16$ as a high-excitation example and $v_i=3$ as a low-excitation example. A vibrational state of $v_i=16$ corresponds to a vibrational energy for NO of 3.34\,eV in the GHM model and 3.4\,eV in  experiment~\cite{kruger2015}. This is close to the dissociation barrier of NO on Au(111) of 3.5\,eV \cite{Falsig2014}. 

\Cref{fig:GardnerMaurerCPA} compares final vibrational distributions for both Markovian and memory-dependent friction. Although the frequency-dependent friction and its conjugate memory kernel in the GHM model are not strictly PSD, the probability distribution from CPA shows that memory-dependent friction dissipates more energy than the strictly PSD Markovian friction, for both the high and low initial states and for both the translational and the vibrational degrees of freedom. This is contrary to the ET model, where memory effects led to a reduction of frictional energy loss. 

In the GHM model, the frequency-dependent friction kernel tensor leads to underdamped dynamics at all geometries except for the $U_0$/$U_1$ state crossing seam. However, the scattering trajectories mostly explore configurations at values of $z$ and $r$ that may be close to the seam, but rarely directly reach it. Even when they reach the crossing seam, this will be close to the reflection point and the velocities will be low, contributing little to the frictional energy loss. In contrast, in the 1D ET model, the friction is most significant close to the barrier, which all trajectories explore. Memory effects lead to an increase of multiquantum energy loss in the final state distributions, leading to a shift of the distributions to lower final vibrational states. We refrain from comparing the results to the final state distributions of previous mixed quantum-classical simulations \cite{gardner2023a} or experiments as we are only calculating energy loss along CPA trajectories and are not considering how memory changes dynamical steering.

Our findings are insensitive to how the memory kernel is averaged. In the SM, Figure S4 shows that local and arithmetic kernel averaging yield minor differences in the memory dissipation overall, vanishing almost entirely in the case of the low initial vibrational state $v_i=3$.

%

\subsection{First-principles memory-dependent electronic friction for NO/Au(111)}

\begin{figure}
    \centering
    \includegraphics[width=\linewidth]{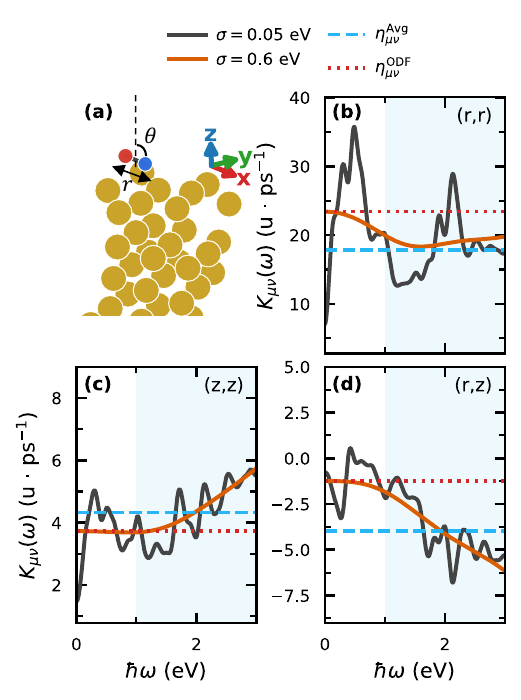}
\caption{(a) NO/Au(111) configuration and coordinate system for trajectory 1 at \(t=219\)~fs (see SM for an overview of the analyzed trajectories). (b--d) Selected components of the mass-weighted friction spectrum tensor, \(K_{\mu\nu}(\omega)\),  shown here for the \((r,r)\), \((z,z)\), and \((r,z)\) components for the geometry shown in panel a. The positive-frequency spectrum obtained from a \(\sigma=0.01\)~eV FHI-aims calculation was additionally Gaussian broadened to total widths of \(\sigma=0.05\) and \(0.6\)~eV. Horizontal lines indicate the Markovian friction values, \(\eta_{\mu\nu}^{\mathrm{ODF}}\) and \(\eta_{\mu\nu}^{\mathrm{Avg}}\) based on the two methods described in Section \ref{sec:CPA}. The shaded region marks the high-energy averaging window used to define \(\eta_{\mu\nu}^{\mathrm{Avg}}\).}
\label{fig:abinitio_frequency_friction}
\end{figure}

With the developed formalism, we can also calculate full-dimensional, tensorial electronic friction kernels from Kohn-Sham DFT, enabling us to revisit the NO/Au(111) scattering problem previously studied with machine learning interatomic potentials and MDEF simulations based on ODF Markov friction tensors \cite{box_2020_determining}. We evaluate DFT frequency-dependent friction tensors along six BOMD trajectories. For these six trajectories, the NO is initialised in a vibrational state of $v_i=16$ and undergoes multiquantum vibrational energy loss. 

In Ref.~\cite{box_2020_determining}, it was established that BOMD underestimates the degree of vibrational energy loss compared to experimental results. MDEF simulations with an ODF friction tensor model were able to improve the description of vibrationally elastic and single quantum inelastic scattering, but otherwise did not improve the description of energy loss beyond BOMD. In this section, we investigate whether the inclusion of memory friction effects will significantly alter the dissipative work done by friction. As we are only evaluating energy loss on BOMD trajectories with CPA, we ignore dynamical steering effects of electronic friction. As a result, we can only report how memory affects the energy loss along a trajectory, not how it affects final vibrational state distributions. However, Box \textit{et al.} \cite{box_2020_determining} approximately accounted for memory effects by scaling the magnitude of Markovian friction by a factor of four, as the friction spectrum showed stronger coupling at higher excitation energies. This brought the simulated final vibrational state distributions much closer to experiment.

As shown in \cref{fig:abinitio_frequency_friction}, the DFT friction spectrum for NO/Au(111) at a representative configuration is strongly anisotropic in the mass-weighted internal-coordinate representation. The magnitude and frequency dependence differ substantially between the stretch component, $\mathcal{K}{rr}$, the surface-normal component, $\mathcal{K}{zz}$, and the stretch--translation coupling, $\mathcal{K}_{rz}$. The diagonal components decrease in magnitude towards the low-frequency limit. They both show a peak at about 0.5 eV. $\mathcal{K}{zz}$ shows a relatively monotonic increase. $\mathcal{K}{rr}$  shows further peaks and modulations at about 2 eV. The magnitude of friction along $r$ is always larger than along the $z$ component, in line with the known anisotropy of electronic friction for this system. However, the ratio of friction between the two coordinates, $r/z$, changes as a function of frequency and ranges between roughly 3 and 7.5. The spectra calculated using a broadening of $\sigma=0.05$~eV retain pronounced structure associated with the underlying electron--hole transition spectrum, whereas the larger broadening of $\sigma=0.6$~eV produces a substantially smoother frequency dependence. 

The behaviour close to $\omega=0$ requires particular care. Within the first-order interband framework used here, the friction vanishes in the combined limits $\sigma\rightarrow0$ and $\omega\rightarrow0$. The low-frequency interband contribution is therefore sensitive to the sampling of low-energy electron--hole transitions and, consequently, to both the applied broadening, the Brillouin zone sampling, and the supercell size. 
This is consistent with the slow convergence of the low-frequency intraband contribution with supercell size reported in Ref.~\cite{box_2020_determining}.  Previously, zero-frequency ODF friction was evaluated using a significant broadening, which provides a numerically converged friction coefficient. Here, $\bm{\eta}^{\mathrm{ODF}}$ is evaluated using $\sigma=0.6$~eV and, as expected, agrees with the zero-frequency limit of the friction spectrum calculated using the same broadening. As shown in \cref{fig:abinitio_frequency_friction}, $\bm{\eta}^{\mathrm{ODF}}$ with small broadenings (black lines) leads to much smaller friction values. $\bm{\eta}^{\mathrm{ODF}}$  therefore sensitively depends on the numerical choice of $\sigma$. 

The finite-frequency averaged Markovian friction, $\bm{\eta}^{\mathrm{Avg}}$ is larger in magnitude than $\bm{\eta}^{\mathrm{ODF}}$ for each of the components shown. This difference follows directly from the frequency dependence of the spectra. Whereas $\bm{\eta}^{\mathrm{ODF}}$ samples the weak linear response in the zero-frequency limit, $\bm{\eta}^{\mathrm{Avg}}$ averages $\bm{\mathcal{K}}(\omega)$ over the (arbitrary) 1--3~eV interval, where the friction spectra have substantially larger magnitudes. The comparison therefore shows that the zero-frequency Markovian friction is considerably weaker than the dissipative coupling sampled over the finite-frequency range relevant to highly vibrationally excited NO. Both Markovian approximations require additional choices in how the frequency-dependent response is reduced to a single effective friction tensor, whereas the non-Markovian treatment retains the full spectrum.

\begin{figure}
    \centering
    \includegraphics[width=\linewidth]{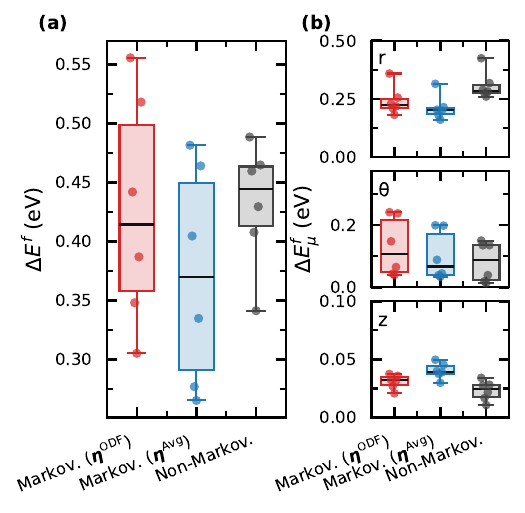}
    \caption{CPA frictional energy losses, $\Delta E^{f}$, evaluated over six representative NO/Au(111) scattering trajectories for three friction models: the Markovian ODF tensor $\boldsymbol{\eta}^{\mathrm{ODF}}$, the Markovian frequency-averaged tensor $\boldsymbol{\eta}^{\mathrm{Avg}}$, and the non-Markovian local memory kernel.  Both $\boldsymbol{\eta}^{\mathrm{Avg}}$ and the non-Markovian kernel are constructed from the $\sigma = 0.05$ eV broadened frequency friction spectrum, using the normalized one-sided cosine transform, an exponential smooth cutoff with $\hbar\omega_{\max}=3.2$ eV. Panel (a): distribution of total final frictional energy losses,  panel (b): the individual energy loss contributions for the internal stretch coordinate $r$, the orientation angle $\theta$ of the molecule, and the centre-of-mass motion along $z$.}
    \label{fig:abinitio_work}
\end{figure}

\cref{fig:abinitio_work} compares the frictional energy dissipation obtained for the considered NO/Au(111) trajectories using three descriptions: the conventional zero-frequency Markovian tensor $\bm{\eta}^{\mathrm{ODF}}$, the finite-frequency averaged Markovian tensor $\bm{\eta}^{\mathrm{Avg}}$, and the endpoint non-Markovian memory kernel defined in \cref{eq:local-phase-point}. The memory treatment provides the most complete description considered here and is therefore used as the reference for assessing the two Markovian approximations. It gives a median energy dissipation of $0.42$~eV across the six trajectories, compared with $0.21$~eV for $\bm{\eta}^{\mathrm{ODF}}$. The zero-frequency ODF treatment therefore substantially underestimates the total energy loss relative to the non-Markovian calculation.

The finite-frequency averaged tensor $\bm{\eta}^{\mathrm{Avg}}$ gives a median total dissipation of $0.37$~eV, which is closer to the full-memory result. This agreement cannot be interpreted as evidence that a single effective Markovian tensor reproduces the underlying dynamics for two reasons. First, $\bm{\eta}^{\mathrm{Avg}}$ is constructed obtained by averaging the friction spectrum over the arbitrary 1--3~eV interval. The optimal energy range will depend on the kinetic energy profile of the trajectories under study. Secondly, as we will show below, memory effects not only change the overall energy loss, but it also differentially affects different modes, affecting the anisotropy of nonadiabatic energy loss.

The importance of accounting for memory effects becomes clear in the mode-resolved energy losses. The median contributions to the total loss associated with $(r,\theta,z)$ are $(0.20, 0.066, 0.039)$~eV for $\bm{\eta}^{\mathrm{Avg}}$, compared with $(0.29, 0.069, 0.017)$~eV for the full memory treatment. Although the two approaches give similar median total dissipation, $\bm{\eta}^{\mathrm{Avg}}$ underestimates the energy loss through the internal stretch coordinate and overestimates the loss along the centre-of-mass motion normal to the surface. Retaining memory therefore redistributes the dissipation towards the high-frequency intramolecular vibration and away from the lower-frequency translational motion. In the case of NO on Au(111), memory effects increase mode anisotropy and significantly enhance vibrational energy loss compared to both the zero-frequency limit and the frequency-averaged Markovian case.

This effect can be understood qualitatively from the structured frequency dependence of the friction spectrum in \Cref{fig:abinitio_frequency_friction}. The N--O stretch probes a higher characteristic frequency range, whereas the centre-of-mass $z$ motion is dominated by lower frequencies. A single uniform average over the 1--3~eV interval can therefore not account for the differential effect of the velocity profile along the trajectory and how it probes different frequencies in the electronic friction kernel. The averaging fails to capture the detailed spectral weight relevant to the stretch motion. The particularly large intramolecular stretch dissipation obtained with the memory kernel likely arises from the pronounced finite-frequency structure in $\mathcal{K}_{rr}(\omega)$ (shown in \Cref{fig:abinitio_frequency_friction}) and the configurational dependence of the friction kernel. Both effects are considered in our non-Markovian treatment via the local memory kernel, as the kernel is updated at each timestep along the trajectory.

The comparison therefore shows that reproducing the median total energy loss is not sufficient to establish the validity of the  Markov approximation. In the present case, $\bm{\eta}^{\mathrm{Avg}}$ gives a similar overall dissipation to the full memory treatment but predicts different partitioning between vibration, rotation, and translation. Since state-resolved scattering observables are directly sensitive to this mode selectivity, the temporal structure of the electronic response that gives rise to the frictional drag on the atoms can remain important even when the total dissipated energy is approximately reproduced by an effective Markovian tensor.

The here presented classical-path analysis evaluates the dissipative work along fixed BOMD trajectories. It does not include feedback of the memory force on the nuclear dynamics and therefore ignores nonadiabatic steering effects that can affect molecule-surface interaction time, trapping probabilities and final energy distributions of scattered projectiles. 

To account for such effects, a fully memory-dependent, tensorial generalised Langevin equation with coloured noise needs to be propagated. Despite the limitations of the current CPA analysis, our results do suggest that the inclusion of memory friction effects may significantly increase the vibrational energy loss observed for NO/Au(111). It was already seen in Ref.~\cite{box_2020_determining} that scaling the diagonal $r$ ODF friction tensor element by a factor of $4$ qualitatively reproduced the experimental vibrational energy loss distribution. For the six trajectories investigated in the limit of ignoring dynamical steering effects, we observe approximately a $ 2.5$-fold increase in the $r$ mode energy dissipation. This is encouraging for the performance of the memory friction method but requires further investigation with a much larger sample of trajectories, for instance by developing machine learning models for the frequency-dependent friction, similar to previously proposed Markovian surrogate models \cite{sachs2025}.

\section{Discussion}

In this work, we have presented a systematic formalism and derivation of memory-dependent electronic friction as a mixed quantum--classical description of nonadiabatic energy exchange between nuclear motion and metallic electrons. 
The formalism is based on a local linearisation of the coupling between electronic excitations and nuclear motion and is applicable to effective independent-electron and mean-field Hamiltonians, including \textit{ab initio} Kohn--Sham electronic structure descriptions. 
Within this framework, the memory effect arises from the fact that the electron-nuclear response depends on the configuration of the atoms and on the strength of the perturbation, i.e., the energy of the excited electrons. The conventional Markovian approximation corresponds to retaining only the static, zero-frequency limit of this response. 
Furthermore, the full configuration-dependent memory kernel arises from a two-time correlation function and generally depends on both the current and past nuclear configurations. This dependence reduces to a local position-dependent kernel whenever the dependence of the electronic Hamiltonian on the nuclear coordinates can be linearised. This result provides a practical route for incorporating memory effects into realistic first-principles electronic friction calculations that we take advantage of in this work. The choice of the configuration within the memory window at which the local friction kernel is evaluated is arbitrary, but we show that choosing the current configuration or an arithmetic mean of the kernel at the current and past time makes virtually no difference for the energy dissipation in the studied systems. 

The first type of Hamiltonian we investigate is the analytically tractable Anderson impurity model, for which we provide a closed-form expression for the memory-dependent electronic friction. This expression recovers the established Markovian result \cite{brandbyge_1995_electronically,jin_practical_2019,gardner2023a,gardner2023} in the zero-frequency limit. This closed-form expression in the linear coupling approximation also exposes a fundamental limitation of electronic friction in this limit: certain combinations of impurity energy level and hybridization strength cause the friction tensor to lose positive semi-definiteness, invalidating the physical picture of electronic friction in a closed system. By deriving the exact threshold for this breakdown, we demonstrate that the applicability of non-Markovian electronic friction in the linearised form is governed not only by the accuracy of the weak coupling approximation that underlies the derivation of electronic friction, but also by the underlying electronic Hamiltonian itself.

The CPA analysis of energy dissipation in the ET model shows that memory effects are strongest when the molecule-surface coupling is weak and, in turn, nonadiabatic coupling is strong and strongly localised. This is also the limit in which Markovian friction has been previously shown to be invalid~\cite{gardner2023a}. Memory effects also consistently yield stronger vibrational relaxation for NO on Au(111), where Markovian friction previously failed to reproduce experiment~\cite{box_2020_determining}, but explicit mixed quantum-classical dynamics based on independent electron surface hopping succeeded~\cite{meng2025}.  While it is clear that the Markov approximation to MDEF is not valid for the study of hyperthermal scattering, it remains unclear whether inclusion of memory is able to (somewhat) extend the realm of applicability of molecular dynamics with electronic friction. Future assessment of this will require high-dimensional dynamic evolution of memory-dependent molecular dynamics with electronic friction that accounts for the tensorial, configuration, and frequency-dependent nature of electronic friction.

The CPA analysis of the full-dimensional \textit{ab initio} trajectories of NO scattering on Au(111) showed that  memory effects alter the partitioning of electronic energy loss and increase mode anisotropy between molecular degrees of freedom. This finding will likely be relevant beyond gas-surface dynamics, as memory effects may also arise in electron-phonon-induced vibrational lifetimes at molecule-metal interfaces and in two-dimensional materials as vibrations of different frequency probe different energy regimes of the electron-phonon response. We hypothesise that memory effects may also affect the temperature dependence of electron-phonon coupling constants for a broad range of materials.

The presented theoretical framework connecting memory-dependent electronic friction across model Hamiltonians and first-principles electronic structure methods will hopefully also help to address other remaining open questions in electronic friction theory, which we have not considered here. This includes the treatment of electron correlation effects~\cite{dou17}, the role of current-induced forces and force contributions that arise from internal and external electromagnetic fields \cite{martinazzo_quantum_2022-1,nazarov_exact-factorization_2026}, and the interplay of quantum nuclear effects and electronic friction~\cite{trenins2025}.


\section*{Supplementary Material}
Supplementary material, containing model parameter details and additional results, can be found below.
\section*{Acknowledgement}
The authors acknowledge support via a UKRI Future Leaders Fellowship [MR/X023109/1], a UKRI Frontier grant [EP/X014088/1], an MSCA postdoctoral fellowship [EP/Z001498/1] and an Alexander-von-Humboldt Professorship. X.L. is supported by EPSRC Doctoral Training Partnership studentships.

Computational resources were provided by the EPSRC-funded HEC Materials Chemistry [EP/L000202/1, EP/R029431/1] and the HPC-CONEXS [EP/X035514/1] consortia for access to the ARCHER2 UK National Computing Service, and the  
EPSRC-funded HPC Midlands+ computing centre for access to Sulis [EP/P020232/1].

C.L.B thanks George Trenins for useful discussions.

\section*{Author declarations}
\subsection*{Conflict of Interest}
The authors have no conflict of interest to declare.

 \section*{Data availability}
The data and the scripts for plotting the figures in this manuscript are available at: \url{https://github.com/maurergroup/memory-friction-paper_figures}.
The code to generate the data will be made publicly accessible during the final step of the publication process. 

\appendix
\section{Imaginary part of retarded electron–hole pair self-energy $\operatorname{Im}\Pi^r_{\mu\nu}$}\label{ap:retarded-electron-hole-pair-self-energy}
The interaction-weighted electron-phonon density of states in matrix form, $\mathbf{\Gamma}$, is related to the electron-phonon self-energy $\tilde{\mathbf{\Pi}}$ through
    \begin{equation}
    \label{eq:pi2gammalü_old}
        \mathbf{\Gamma}(\omega\,;\mathbf{x}) = \frac{1}{2\pi i}  
        \mathcal{F}\{\tilde{\mathbf{\Pi}}(t-t'\,;\mathbf{x})\},
    \end{equation}
with Fourier transformation operator $\mathcal{F}$. The ($\mu$,$\nu$) mode component of $\tilde{\mathbf{\Pi}}$ is given by the difference between its retarded, $\Pi^{r}_{\mu\nu}$, and advanced, $\Pi^{a}_{\mu\nu}$ self-energy components,
    \begin{equation}
    \tilde{\Pi}_{\mu\nu} = \Pi_{\mu\nu}^{r}
    - \Pi_{\mu\nu}^{a} = \Pi_{\mu\nu}^{r} - \left[\Pi_{\nu\mu}^{r}\right]^*,
    \end{equation}
    where the relation $\Pi^a_{\mu\nu}(\omega\,;\mathbf{x}) = \left[\Pi^{r}_{\nu\mu}(\omega\,;\mathbf{x})\right]^*$ follows from the Hermiticity of the Hamiltonian.
    To extract the real part, we write $\Pi^{r}_{\mu\nu} = A + Bi$ and $\Pi^{r}_{\nu\mu} = C + Di$ with $A, B, C, D \in \mathbb{R}^{n_I\times n_I}$. A straightforward calculation then gives
    \begin{equation*}
        \Gamma_{\mu\nu} = \frac{1}{2\pi} \left[(B+D) - i(A-C)\right],
    \end{equation*}
    from which we read off the real part that gives us \cref{eq:ReGammaLu}.

Denoting $\tau = t - t'$ and separating \cref{eq:_self_energy} into two terms, $T_1(\tau\,;\mathbf{x})$ and $T_2(\tau\,;\mathbf{x})$ can be treated in turn. 
    The first term reads
\begin{equation}
\label{eq:_self_energy_term1}
    T_1(\tau\,;\mathbf{x}) = -i\hbar \, \operatorname{Tr} \left[ \mathcal{M}^\mu G_0^{r}(\tau\,;\mathbf{x}) \, \mathcal{M}^\nu G_0^{<}(-\tau\,;\mathbf{x}) \right].
\end{equation}
To transform this into frequency space, we first express the non-interacting retarded and lesser Green's functions in terms of their energy-space representations:
\begin{align}
    G_{0}^{r}(\tau\,;\mathbf{x}) &= \int \frac{d\epsilon_{1}}{2\pi\hbar} \, e^{-i\epsilon_{1}\tau/\hbar} \, G_{0}^{r}(\epsilon\,;\mathbf{x}), \\
    G_{0}^{<}(-\tau\,;\mathbf{x}) &= \int \frac{d\epsilon_{2}}{2\pi\hbar} \, e^{i\epsilon_{2}\tau/\hbar} \, G_{0}^{<}(\epsilon\,;\mathbf{x}).
\end{align}

Substituting these into \cref{eq:_self_energy_term1}, we obtain
\begin{equation}
\begin{split}
T_{1}(\tau\,;\mathbf{x}) = {} &
-i\hbar \int \frac{d\epsilon_{1}}{2\pi\hbar} 
\int \frac{d\epsilon_{2}}{2\pi\hbar} \,
e^{-i(\epsilon_{1}-\epsilon_{2})\tau/\hbar} \\
& \times \operatorname{Tr} \left[ 
\mathcal{M}^\mu G_{0}^{r}(\epsilon\,;\mathbf{x}) \,
M^{\nu} G_{0}^{<}(\epsilon\,;\mathbf{x}) 
\right].
\end{split}
\end{equation}
We now take the Fourier transform of $T_1(\tau)$:
\begin{equation}
\label{eq:T_1_frequency_time}
    T_1(\omega\,;\mathbf{x}) = \int d\tau \, e^{i\omega\tau} \, T_1(\tau\,;\mathbf{x}).
\end{equation}
To evaluate this, we make use of the identity
\begin{equation*}
    \int d\tau \, e^{i[\omega - (\epsilon_1 - \epsilon_2)/\hbar]\tau} = 2\pi \, \delta\!\left(\omega - \frac{\epsilon_1 - \epsilon_2}{\hbar}\right),
\end{equation*}
which gives
\begin{equation}
\label{eq:T_frequency}
\begin{split}
T_1(\omega\,;\mathbf{x}) = {} &
-i\hbar \int \frac{d\epsilon_1}{2\pi\hbar} 
\int \frac{d\epsilon_2}{2\pi\hbar} \,
2\pi \, \delta\!\left(\omega - \frac{\epsilon_1 - \epsilon_2}{\hbar}\right) \\
& \times \operatorname{Tr} \left[ 
\mathcal{M}^\mu G_0^{r}(\epsilon\,;\mathbf{x}) \,
\mathcal{M}^\nu G_0^{<}(\epsilon\,;\mathbf{x}) 
\right].
\end{split}
\end{equation}
Using the rescaling property of the delta function,
\begin{equation*}
    \delta\!\left(\omega - \frac{\epsilon_1 - \epsilon_2}{\hbar}\right) = \hbar \, \delta(\epsilon_1 - \epsilon_2 - \hbar\omega),
\end{equation*}
and integrating out $\epsilon_1$, \cref{eq:T_frequency} reduces to
\begin{equation}
\label{eq:_self_energy_T1}
    T_1(\omega\,;\mathbf{x}) = -i\hbar \int \frac{d\epsilon}{2\pi\hbar} \, \operatorname{Tr} \left[ \mathcal{M}^\mu G_0^{r}(\epsilon+ \hbar\omega\,;\mathbf{x}) \, \mathcal{M}^\nu G_0^{<}(\epsilon\,;\mathbf{x}) \right].
 \end{equation}
Repeating these steps for the second term, we obtain
\begin{equation}
\label{eq:_self_energy_T2}
T_2(\omega\,;\mathbf{x}) = - i \hbar \int \frac{d\epsilon}{2\pi \hbar} \text{Tr} \left[ \mathcal{M}^\mu G_0^< (\epsilon + \hbar \omega\,;\mathbf{x}) \mathcal{M}^\nu G_0^a(\epsilon\,;\mathbf{x}) \right].    
\end{equation}
At electronic equilibrium, the lesser Green's function 
satisfies
\begin{equation}
    G^<_0(\epsilon\,;\mathbf{x}) = n_F(\epsilon)
    \left[G^a_0(\epsilon\,;\mathbf{x}) - G^r_0(\epsilon\,;\mathbf{x})\right],
\end{equation}
where $G^a_0 = [G^r_0]^\dagger$. Any matrix can be 
decomposed into its Hermitian and anti-Hermitian parts. 
For the retarded Green's function, we define
\begin{equation}
\begin{split}
\operatorname{Re}[G^r_0] \equiv {} &
\frac{G^r_0 + [G^r_0]^\dagger}{2} 
= \frac{G^r_0 + G^a_0}{2}, \\
\operatorname{Im}[G^r_0] \equiv {} &
\frac{G^r_0 - [G^r_0]^\dagger}{2i} 
= \frac{G^r_0 - G^a_0}{2i}.
\end{split}
\end{equation}
where both $\operatorname{Re}[G^r_0]$ and 
$\operatorname{Im}[G^r_0]$ are Hermitian matrices. 
Note that for matrices, these denote the Hermitian 
decomposition, not element-wise real and imaginary 
parts. The retarded and advanced Green's functions 
are then expressed as
\begin{equation}
\begin{split}
G^r_0 = {} &
\operatorname{Re}[G^r_0] 
+ i\,\operatorname{Im}[G^r_0], \\
G^a_0 = {} &
\operatorname{Re}[G^r_0] 
- i\,\operatorname{Im}[G^r_0].
\end{split}
\end{equation}

Defining the spectral function matrix
\begin{equation}
    \mathcal{A}(\epsilon\,;\mathbf{x}) = i\left[G^r_0(\epsilon\,;\mathbf{x}) 
    - G^a_0(\epsilon\,;\mathbf{x})\right] 
    = -2\,\operatorname{Im}[G^r_0(\epsilon\,;\mathbf{x})],
\end{equation}
the lesser Green's function takes the compact form
\begin{equation}
    G^<_0(\epsilon\,;\mathbf{x}) = -i\, n_F(\epsilon)\, 
    \mathcal{A}(\epsilon\,;\mathbf{x}).
\end{equation}
Substituting into $T_1(\omega\,;\mathbf{x})$ and $T_2(\omega\,;\mathbf{x})$ from 
\cref{eq:_self_energy_T1,eq:_self_energy_T2}, 
and decomposing $G^r_0$ and $G^a_0$ into their Hermitian 
and anti-Hermitian parts, we collect the imaginary 
contributions following the same procedure as in the 
scalar case detailed earlier. The result is 
    \begin{equation}
    \begin{split}
    \operatorname{Im}\Pi^r_{\mu\nu}(\omega\,;\mathbf{x}) = {} &
    \int \frac{d\epsilon}{2\pi}\;
    \operatorname{Tr}\left[\mathcal{M}^\mu\,\mathcal{A}
    (\epsilon+\hbar\omega\,;\mathbf{x})\, \mathcal{M}^\nu\,
    \mathcal{A}(\epsilon\,;\mathbf{x})\right] \\
    & \times \left[n_F(\epsilon+\hbar\omega)
    - n_F(\epsilon)\right],
    \end{split}
    \end{equation}
which is \cref{eq:imaginaryPi_matrix}.

\section{From single mode to the multi-mode coupling friction 
matrix}\label{ap:multi-orbital-to-single-energy-level}
Brandbyge \textit{et al.} \cite{brandbyge_1995_electronically} derived 
the Markovian electronic friction coefficient for a single adsorbate 
mode~$x$ within the Newns--Anderson model:

\begin{equation}
    \eta(x) =  \frac{1}{\pi}\int_{-\infty}^{\infty} \mathrm{d}\epsilon \left(\frac{\partial \delta}{\partial x}(x\,;\epsilon)\right)^2 \left(-\frac{\mathrm{d} n_F}{\mathrm{d}\epsilon}(\epsilon)\right)
\end{equation}
where $n_F$ stands for the Fermi-Dirac distribution and the scattering 
phase shift is 
\begin{equation}
    \delta(x\,;\epsilon) = \frac{\pi}{2} -\arctan \left(\frac{h(x) + \mathcal{H}(\epsilon)-\epsilon}{\Delta(x\,;\epsilon)}\right)
\end{equation}
Differentiating with respect to $x$ and recognizing the adsorbate 
spectral function 
$\mathcal{A}(x\,;\epsilon) = 2\Delta/[\Delta^2 + (h + \mathcal{H} 
- \epsilon)^2]$, the phase-shift derivative becomes
\begin{equation}
    \frac{\partial\delta}{\partial x}
    = -\frac{\mathcal{A}}{2}
    \left(
        \frac{\partial(h + \mathcal{H})}{\partial x}
        + \frac{\epsilon - h - \mathcal{H}}{\Delta}\,
          \frac{\partial\Delta}{\partial x}
    \right).
\end{equation}
The bracket is precisely the 
electron--nuclear coupling $\mathcal{M}^{\mu}$ defined in 
\cref{eq:Mmatrix,eq:Ak_old}, so that 
$\partial\delta/\partial x_\mu = -\mathcal{A}\,\mathcal{M}^{\mu}/2$. 
Substituting, Brandbyge's Markovian friction takes the form
\begin{equation}
    \eta(x) = \frac{1}{4\pi}\int_{-\infty}^{\infty} 
    \mathrm{d}\epsilon\;
    \mathcal{A}(x\,;\epsilon)^2\;
    \bigl[\mathcal{M}(x\,;\epsilon)\bigr]^2
    \left(-\frac{\mathrm{d} n_F}{\mathrm{d}\epsilon}\right),
\end{equation}
which is the single-orbital, single-mode reduction of the general 
multi-orbital Markovian friction tensor 
$ \mathcal{K}(\omega =0,\mathbf{x}) \propto 
\mathrm{Tr}[\mathcal{M}^{\mu}A\,\mathcal{M}^{\nu}A]\,(-\mathrm{d}n_F/\mathrm{d}\epsilon)$ 
obtained from the adiabatic limit of L\"u \textit{et al.}'s 
formalism~\cite{Luu2012}. In the present work, we generalize to 
multiple nuclear degrees of freedom; in the single-resonant-level 
limit of the NAH, 
$\mathcal{M}^{\mu}$ takes the explicit form
\begin{equation}
\begin{aligned}
    \mathcal{M}^{\mu}(\epsilon\,;\mathbf{x}) 
    = {} & \frac{\partial\bigl(h(\mathbf{x}) 
          + \mathcal{H}(\epsilon\,;\mathbf{x})\bigr)}{\partial x_\mu} \\
      & + \frac{\epsilon - h(\mathbf{x}) 
          - \mathcal{H}(\epsilon\,;\mathbf{x})}
         {\Delta(\epsilon\,;\mathbf{x})}\,
         \frac{\partial\Delta(\epsilon\,;\mathbf{x})}{\partial x_\mu}.
\end{aligned}
\end{equation}

\section{Positive semi-definiteness of memory electronic friction in NAH}\label{app:PSD}
The frequency-dependent friction, $\mathcal{K}_{\mu\nu}(\omega\,;\mathbf{x}) = \Lambda_{\mu\nu}(\omega\,;\mathbf{x})/2$, discussed in \cref{sec:Single position based friction kernel,sec:NAH} is defined as
\begin{equation}\label{eq:Lambda}
    \Lambda_{\mu\nu}(\omega\,; \mathbf{x}) = -\frac{1}{2\omega} \left[\operatorname{Im}\Pi^r_{\mu\nu}(\omega\,; \mathbf{x}) + \operatorname{Im}\Pi^r_{\nu\mu}(\omega\,; \mathbf{x})\right],
\end{equation}
where
\begin{equation}
\label{eq:Ikl_split1}
\begin{split}
    \operatorname{Im}\Pi^r_{\mu\nu}(\omega\,; \mathbf{x}) = \int^{\infty}_{-\infty}\frac{\mathrm{d}\epsilon}{2\pi} \, & A^\mu(\epsilon\,;\mathbf{x})\; A^\nu(\hbar\omega+\epsilon\,;\mathbf{x}) \\
    & \times \bigl[n_F(\hbar\omega+\epsilon) - n_F(\epsilon)\bigr],
\end{split}
\end{equation}
with the vertex function
\begin{equation}\label{eq:Ak}
    A^\mu(\epsilon\,;\mathbf{x}) = \mathcal{A}(\epsilon\,;\mathbf{x}) \left[ \frac{\partial h(\mathbf{x})}{\partial x_\mu}
    + \frac{\epsilon - h(\mathbf{x})}{\Delta(\epsilon\,;\mathbf{x})} \frac{\partial \Delta(\epsilon\,;\mathbf{x})}{\partial x_\mu} \right].
\end{equation}
Here $\mathcal{A}(\epsilon\,;\mathbf{x})$ is the spectral density, $h(\mathbf{x})$ the electronic on-site energy, $\Delta(\epsilon\,;\mathbf{x})$ the hybridization function, and $n_F(\epsilon)=[\mathrm{e}^{\beta\epsilon}+1]^{-1}$ the Fermi--Dirac distribution as discussed in the main text.

For any real vector $\mathbf{v}=(v_1,\dots,v_N)^\top$, define the scalar function
\begin{equation}\label{eq:Av}
    \bar{A}(\epsilon) \;\equiv\; \sum_\mu v_\mu\, A^\mu(\epsilon\,;\mathbf{x}).
\end{equation}
Then the symmetrised combination gives
\begin{equation}\label{eq:quadform}
\begin{split}
    \mathbf{v}^\top \mathbf{\Lambda}(\omega\,; \mathbf{x})\,\mathbf{v} 
    = {} & -\frac{1}{\omega}\int_{-\infty}^{\infty}\frac{\mathrm{d}\epsilon}{2\pi}\; \bar{A}(\epsilon)\,\bar{A}(\hbar\omega+\epsilon) \\
    & \times \bigl[n_F(\hbar\omega+\epsilon)-n_F(\epsilon)\bigr].
\end{split}
\end{equation}

Using the identity
\begin{equation}\label{eq:db}
    n_F(\hbar\omega+\epsilon) - n_F(\epsilon)
    = -\bigl(\mathrm{e}^{\beta\hbar\omega}-1\bigr)\,n_F(\hbar\omega+\epsilon)\bigl[1-n_F(\epsilon)\bigr],
\end{equation}
we rewrite the quadratic form \cref{eq:quadform} as
\begin{equation}
\label{eq:quadform_final}
\begin{split}
\mathbf{v}^\top\mathbf{\Lambda}(\omega\,; \mathbf{x})\,\mathbf{v}
&= \frac{\mathrm{e}^{\beta\hbar\omega}-1}{\omega}
   \int_{-\infty}^{\infty}\frac{\mathrm{d}\epsilon}{2\pi}\;
   \bar{A}(\epsilon)\,\bar{A}(\hbar\omega+\epsilon) \\
&\qquad\times
   n_F(\hbar\omega+\epsilon)
   \bigl[1-n_F(\epsilon)\bigr].
\end{split}
\end{equation}
The prefactor $(\mathrm{e}^{\beta\hbar\omega}-1)/\omega > 0$ for all $\omega\neq 0$, and the Fermi phase-space product $n_F(\hbar\omega+\epsilon)[1-n_F(\epsilon)]\geq 0$.  Therefore, the general PSD condition in \cref{eq:quadform_final} holds for any $\omega$. At the zero-frequency limit or the well-known Markovian electronic friction, the PSD of the friction matrix always holds since
\begin{equation}
\label{eq:omega0}
    \lim_{\omega\rightarrow0}\mathbf{v}^\top\mathbf{\Lambda}(\omega\,; \mathbf{x})\,\mathbf{v}
    = -\frac{\hbar}{2\pi}\int_{-\infty}^{\infty}\mathrm{d}\epsilon\;\bar{A}(\epsilon)^2\,\dfrac{\mathrm{d} n_F(\epsilon)}{\mathrm{d}\epsilon}\;\geq\; 0
\end{equation}
is guaranteed.

In the case that the spectral density is a positive scalar $\mathcal{A}$, the vertex function simplifies to 
\begin{equation}\label{eq:Ak_WBL}
    A^\mu(\epsilon\,;\mathbf{x}) = \mathcal{A}\left[\frac{\partial h}{\partial x_\mu}
    + \frac{\epsilon - h}{\Delta}\,\frac{\partial\Delta}{\partial x_\mu}\right]
    = \mathcal{A}\bigl(c_\mu + d_\mu\,\epsilon\bigr),
\end{equation}
where the coefficients are
\begin{equation}\label{eq:ck_dk}
    c_\mu \;\equiv\; \frac{\partial h}{\partial x_\mu} - \frac{h}{\Delta}\,\frac{\partial\Delta}{\partial x_\mu},
    \qquad
    d_\mu \;\equiv\; \frac{1}{\Delta}\,\frac{\partial\Delta}{\partial x_\mu}.
\end{equation}
with their conjugate vector $\mathbf{c}=(c_1,\dots,c_N)^\top$ and $\mathbf{d}=(d_1,\dots,d_N)^\top$.
Using \cref{eq:Av}, the projected vertex function can then be written as
\begin{equation}
\label{eq:A_bar}
\bar{A}(\epsilon)
= \mathcal{A}\bigl(\bar{c}+\bar{d}\epsilon\bigr),
\end{equation}
with
\begin{equation}
\bar{c}
= \sum_\mu v_\mu c_\mu
= \mathbf{v}^{\top}\mathbf{c},
\qquad
\bar{d}
= \sum_\mu v_\mu d_\mu
= \mathbf{v}^{\top}\mathbf{d}.
\end{equation}
The product of two projected vertices in \cref{eq:quadform} therefore becomes
\begin{equation}
\begin{aligned}
\bar{A}(\epsilon)
\bar{A}(\hbar\omega+\epsilon)
= \mathcal{A}^2 \Big[
\bar{c}^2
+ \bar{c}\bar{d}
(\hbar\omega + 2\epsilon)
+ \bar{d}^2
\epsilon(\hbar\omega+\epsilon)
\Big].
\end{aligned}
\end{equation}
\subsection*{Zero-temperature limit evaluation at single degree of freedom}
At $T=0$, $n_F(\epsilon)=\theta(-\epsilon)$, and for $\omega>0$ the Fermi window restricts $\epsilon$ to
\begin{equation}
    \epsilon \in (-\hbar\omega,\,0),
\end{equation}
within which $n_F(\hbar\omega+\epsilon)-n_F(\epsilon) = -1$.
The elementary integrals over $\epsilon\in(-\hbar\omega,0)$ assemble the result:
\begin{equation}\label{eq:integral_result}
    \int_{-\hbar\omega}^{0}\mathrm{d}\epsilon\;\bar{A}(\epsilon)\,\bar{A}(\hbar\omega+\epsilon)
    = \mathcal{A}^2\,\hbar\omega\left(\bar{c}^2 - \frac{(\hbar\omega)^2}{6}\,\bar{d}^2\right).
\end{equation}

In the case of a single degree of freedom, $\bar{c}$ and $\bar{d}$ simplify to $v_1 c$ and $v_1 d$ respectively with
\begin{equation}
    c = \frac{\partial h}{\partial x} - \frac{h}{\Delta}\,\frac{\partial\Delta}{\partial x},
    \qquad
    d = \frac{1}{\Delta}\,\frac{\partial\Delta}{\partial x}.
\end{equation}
Then, electronic friction PSD definition \cref{eq:quadform} becomes
\begin{align}
    \mathbf{v}^\top\mathbf{\Lambda}(\omega\,;\mathbf{x})\,\mathbf{v}\Big|_{T=0}
    &= -\frac{1}{\omega}\cdot\frac{1}{2\pi}\cdot(-1)\cdot \mathcal{A}^2\,\hbar\omega\, v_1^2 \nonumber\\
    &\quad\times \left(c^2 - \frac{(\hbar\omega)^2}{6}\,d^2\right) \nonumber\\
    &= \frac{\mathcal{A}^2\,\hbar v_1^2}{2\pi}\left(c^2 - \frac{(\hbar\omega)^2}{6}\,d^2\right).
\end{align}

Hence, it yields the impurity-bath coupling part in the threshold frequency \cref{eq:threshold_frequency}
\begin{equation}
\label{eq:threshold-frequency-0}
    \omega_{*}^{(T=0)} = \frac{\sqrt{6}}{\hbar}\left|\frac{\partial_x h}{\partial_x \Delta}\Delta - h\right|.
\end{equation}

\subsection*{Finite-temperature thermal correction to the threshold
frequency $\omega_*$ at single degree
of freedom}
At finite temperature the Fermi-window integrals in \cref{eq:quadform}
can be solved in a closd form. 
\begin{equation}\label{eq:In_def}
\begin{aligned}
    I_n(a) &\equiv \int_{-\infty}^{\infty}\mathrm{d}\epsilon\;\epsilon^{\,n}
    \bigl[n_F(a+\epsilon)-n_F(\epsilon)\bigr], \\
    &\quad n = 0,1,2, \quad a \equiv \hbar \omega
\end{aligned}
\end{equation}
and the Sommerfeld moments~\cite{Sommerfeld1928}
\begin{equation}\label{eq:moments}
\begin{split}
    \mu_n &\equiv -\int_{-\infty}^{\infty}\mathrm{d}\epsilon\;
    \epsilon^{\,n}\,n_F'(\epsilon), \\
    &\quad \mu_0 = 1,\quad \mu_1 = 0,\quad \mu_2 = \frac{\pi^2}{3}(k_\mathrm{B}T)^2 .
\end{split}
\end{equation}
Since $\bar A(\epsilon) = \mathcal{A}(\bar c + \bar d\,\epsilon)$ is linear, the
integrand of \cref{eq:quadform} is quadratic and the moment expansion
terminates at $\mu_2$.

With $I_n(0) = 0$ and
$$I_n'(a) \equiv \frac{\mathrm{d} I_n}{\mathrm{d} a} = \int^{\infty}_{-\infty}\mathrm{d}u\,(u-a)^n n_F'(u),$$
giving $I_0'(a) = -1$, $I_1'(a) = a$, $I_2'(a) = -\mu_2 - a^2$, hence
\begin{equation}\label{eq:In_closed}
    I_0(a) = -a,
    \qquad
    I_1(a) = \frac{(a)^{2}}{2},
    \qquad
    I_2(a) = -\mu_2\,a - \frac{(a)^{3}}{3}.
\end{equation}
Inserting
$\bar A(\epsilon)\bar A(\hbar\omega+\epsilon)
 = \mathcal{A}^2[(\bar c^{\,2}+\bar c\bar d\,\hbar\omega)
 + (2\bar c\bar d+\bar d^{\,2}\hbar\omega)\epsilon
 + \bar d^{\,2}\epsilon^{2}]$
into \cref{eq:quadform}, the cross term cancels 
($\hbar\omega I_0 + 2I_1 = 0$; note $I_1$ is $T$-independent), leaving
\begin{equation}\label{eq:quadform_finiteT}
    \mathbf{v}^{\top}\mathbf{\Lambda}(\omega\,;\mathbf{x})\,\mathbf{v}
    = \frac{\mathcal{A}^2\,\hbar}{2\pi}
    \left[\bar c^{\,2}
    + \left(\frac{\pi^2 (k_\mathrm{B}T)^2}{3}
    - \frac{(\hbar\omega)^{2}}{6}\right)\bar d^{\,2}\right],
\end{equation}
exact at a finite $T$.
The threshold frequency $\omega_*$ for a single degree of freedom at a finite temperature $T$ is given as 
\begin{equation}
\label{eq:threshold-frequency-T}
    \omega_* = \sqrt{(\omega_{*}^{(T=0)})^{2} + \frac{2\pi^{2}(k_{B}T)^{2}}{\hbar^{2}}}
\end{equation}
Here, $\frac{\sqrt{2}\pi k_{\mathrm{B}} T}{\hbar}$ represents the finite temperature correction to the PSD threshold frequency. 

\subsection*{Memory friction matrix PSD condition}

Since $\bar{c}^2=\mathbf{v}^{\top}(\mathbf{c}\mathbf{c}^{\top})\mathbf{v}$ and $\bar{d}^2=\mathbf{v}^{\top}(\mathbf{d}\mathbf{d}^{\top})\mathbf{v}$ hold for every $\mathbf{v}\in\mathbb{R}^N$, we identify the friction matrix at zero and finite temperature as
\begin{align} 
\label{eq:Lambda_matrix_T0} 
\mathbf{\Lambda}(\omega\,; \mathbf{x})\Big|_{T=0} &= \frac{\mathcal{A}^2\,\hbar}{2\pi}\left[\mathbf{c}\,\mathbf{c}^{\top} - \frac{(\hbar\omega)^2}{6}\;\mathbf{d}\,\mathbf{d}^{\top}\right], \\ 
\label{eq:Lambda_matrix_finiteT} 
\begin{split}
\mathbf{\Lambda}(\omega\,; \mathbf{x})\Big|_{T>0} &= \frac{\mathcal{A}^2\,\hbar}{2\pi}\left[\mathbf{c}\,\mathbf{c}^{\top} + \left(\frac{\pi^2(k_{\mathrm{B}}T)^2}{3} - \frac{(\hbar\omega)^2}{6}\right)\mathbf{d}\,\mathbf{d}^{\top}\right] \\
&\qquad + O\bigl((\hbar\omega)^2(k_{\mathrm{B}}T)^2\bigr). 
\end{split}
\end{align}

When $\mathbf{c}\parallel\mathbf{d}$ (i.e.\ $\mathbf{c}=\lambda\mathbf{d}$ for some scalar $\lambda$), both expressions reduce to a rank-1 matrix proportional to $\mathbf{d}\mathbf{d}^{\top}$, recovering the single degree of freedom threshold
\cref{eq:threshold-frequency-T} for its sole non-zero eigenmode.  This collinearity condition is equivalent to $\nabla h\parallel\nabla\Delta$:
\begin{equation}\label{eq:collinear_condition}
    \frac{\partial h/\partial x_1}{\partial\Delta/\partial x_1}
    = \frac{\partial h/\partial x_2}{\partial\Delta/\partial x_2}
    = \cdots
    = \frac{\partial h/\partial x_N}{\partial\Delta/\partial x_N}.
\end{equation}

When $\mathbf{c}\not\parallel\mathbf{d}$, the situation changes.  At $T=0$, $\mathbf{\Lambda}(0)=\frac{\mathcal{A}^2\hbar}{2\pi}\,\mathbf{c}\mathbf{c}^{\top}$ is rank-1 with a zero eigenvalue, and for $N=2$ the determinant of \cref{eq:Lambda_matrix_T0} is
\begin{equation}\label{eq:det_T0}
    \det\mathbf{\Lambda}(\omega\,; \mathbf{x})\Big|_{T=0}
    = -\frac{\mathcal{A}^4\,\hbar^2}{24\pi^2}\;(\hbar\omega)^2\;(c_1 d_2 - c_2 d_1)^2
    \;\leq\; 0,
\end{equation}
forcing one eigenvalue to be negative at every $\omega\neq 0$.  At finite temperature, however, the thermal correction promotes $\mathbf{\Lambda}(0)$ to full rank.  The coefficient of $\mathbf{d}\mathbf{d}^{\top}$ in \cref{eq:Lambda_matrix_finiteT} remains non-negative for $|\omega|\leq\omega_c$, where
\begin{equation}\label{eq:omega_c_multi}
    \omega_c = \frac{\pi\sqrt{2}\,k_{\mathrm{B}}T}{\hbar},
\end{equation}
guaranteeing PSD regardless of the angle between $\nabla h$ and $\nabla\Delta$.



%

\end{document}


\title[]{Supplemental Material for \\ "Memory-dependent electronic friction for nonadiabatic dynamics at metal surfaces"}

\author{Xuexun Lu} 
\affiliation{Department of Chemistry, University of Warwick, Coventry, CV4 7AL, United Kingdom}

\author{Connor L. Box}
\affiliation{Yusuf Hamied Department of Chemistry, University of Cambridge, CB2 1EW, Cambridge, United Kingdom}

\author{Nils Hertl}
\affiliation{Department of Chemistry, University of Warwick, Coventry, CV4 7AL, United Kingdom}
\affiliation{Department of Physics, University of Warwick, Coventry, CV4 7AL, United Kingdom}

\author{Reinhard J. Maurer}
\email{reinhard.maurer@univie.ac.at}
\affiliation{Department of Chemistry, University of Warwick, Coventry, CV4 7AL, United Kingdom}
\affiliation{University of Vienna, Faculty of Physics, University of Vienna, Kolingasse 14{--}16, 1090 Vienna, Austria}
\affiliation{Institute of Physical Chemistry, Georg-August University, Göttingen 37077, Germany}
\affiliation{Max-Planck-Institute for Multidisciplinary Sciences, Göttingen 37077, Germany}

\keywords{}

\maketitle

\clearpage
\setcounter{table}{0}
\section{Model Hamiltionian parameter tables}

\begin{table}[htbp]
\caption{Parameters for the Erpenbeck--Thoss model~\cite{ErpenbeckThoss,GardnerEPT}.\label{tab:ET_parameters_SM}}
\begin{ruledtabular}
\begin{tabular}{@{}lS[table-format=-1.2]l@{\hspace{1.5em}}lS[table-format=2.2]l@{\hspace{1.5em}}lS[table-format=-2.4]l@{}}
$D_e$      & 3.52  & \si{\eV}            & $x_0$      & 1.78  & \si{\angstrom}      & $a$        & 1.7361 & \si{\angstrom^{-1}} \\
$D_1$      & 4.52  & \si{\eV}            & $D_2$      & 0.79  & \si{\eV}            & $a'$       & 1.379  & \si{\angstrom^{-1}} \\
$V_\infty$ & -1.5  & \si{\eV}            & $q$        & 0.05  & {1}                 & $\tilde a$ & 0.5    & \si{\angstrom}      \\
$\tilde x$ & 3.5   & \si{\angstrom}      & $m$        & 10.54 & \si{u}              & $c$        & -45.7  & \si{\milli\eV}      \\
\end{tabular}
\end{ruledtabular}
\end{table}

  \begin{table*}[htbp]
  \caption{Parameters for the Gardner--Habershon--Maurer model \cite{gardner2023a} for
  NO/Au(111).\label{tab:NOAu_parameters_SM}}
  \begin{ruledtabular}
  \begin{tabular}{@{}lS[table-format=1.4]l@{\hspace{1.5em}}lS[table-format=2.4]l
  @{\hspace{1.5em}}lS[table-format=2.4]l@{}}
  \multicolumn{3}{@{}c}{\textbf{NO Morse}} &
  \multicolumn{3}{c}{\textbf{Hybridization}} &
  \multicolumn{3}{c@{}}{\textbf{NO/Au}} \\
  \hline
  $r_0$ & 1.1510  & \si{\angstrom}   & $\Delta_0$   & 0.75    & \si{\eV}         &
  $b_0$ & 1.9535 & \si{\angstrom^{-1}} \\
  $a_0$ & 2.7968  & \si{\angstrom^{-1}} & $\tilde a$     & 10 & \si{\angstrom}
  & $z_0$ & -0.26876 & \si{\angstrom} \\
  $D_0$ & 6.610   & \si{\eV}         &  &      &    &
  $c_0$ & 6.5713 & \si{\eV} \\
  $\bar{D}_0$    & 27.2114 &   \si{\eV}           &            &        &                  &
  $a_1$ & 2.5194 & \si{\angstrom^{-1}} \\
        &         &                  &            &        &                  &
  $r_1$ & 1.2950 & \si{\angstrom} \\
        &         &                  &            &        &                  &
  $D_1$ & 4.1528 & \si{\eV} \\
        &         &                  &            &        &                  &
  $a_2$ & 1.0015 & \si{\angstrom^{-1}} \\
        &         &                  &            &        &                  &
  $z_1$ & 1.2350 & \si{\angstrom} \\
        &         &                  &            &        &                  &
  $D_2$ & 2.4171 & \si{\eV} \\
        &         &                  &            &        &                  &
  $c_1$ & 8.9587 & \si{\eV} \\
  \end{tabular}
  \end{ruledtabular}
  \end{table*}

\section{Frequency-dependent friction from Anderson Hamiltonian}

\begin{figure}[H]
    \centering
    \includegraphics[width=1.0\linewidth]{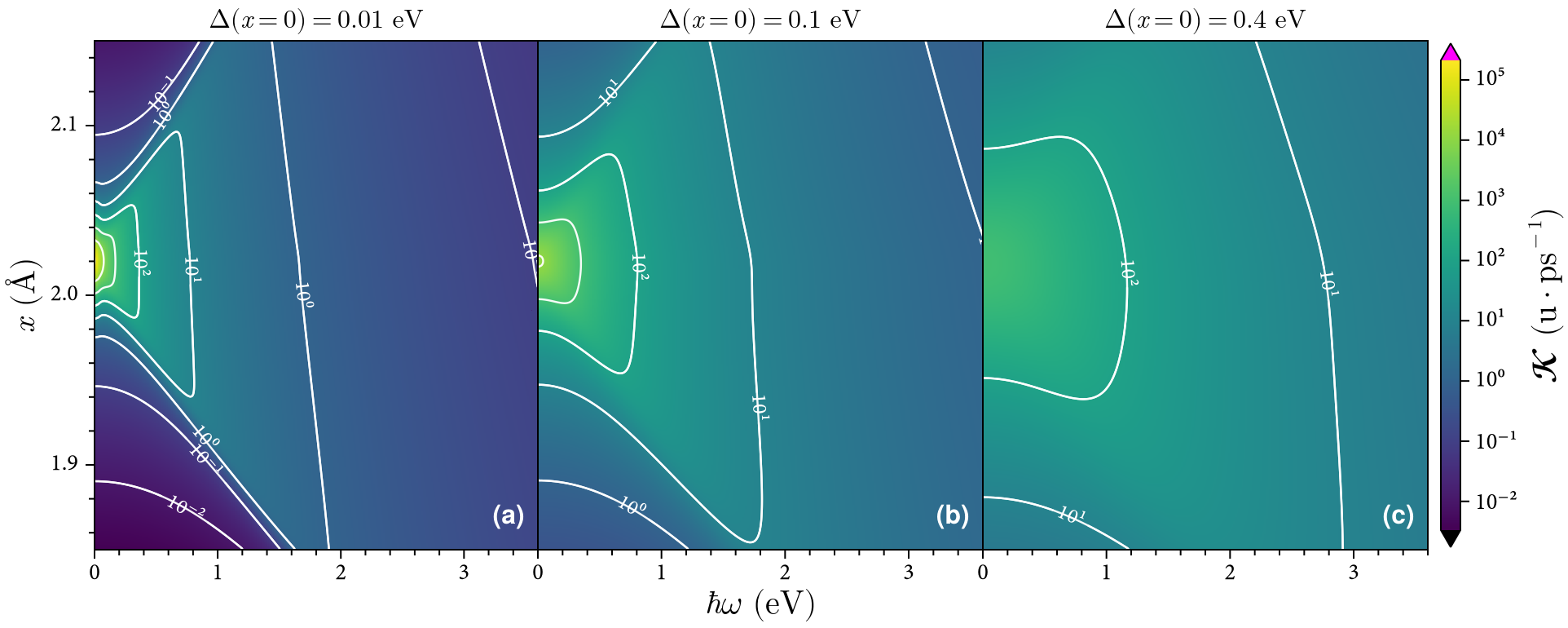}
    \caption{\textbf{Frequency-dependent electronic friction for different values of hybridization
strength.} Two-dimensional map of the frequency-dependent friction
$\mathcal{K}(x,\omega)$ (color, logarithmic scale, in $\mathrm{u\cdot ps^{-1}}$)
for the Erpenbeck--Thoss model~\cite{ErpenbeckThoss,GardnerEPT}, as a function of nuclear
coordinate $x$ and energy $\hbar\omega$, at temperature $T = 300\,\mathrm{K}$.
Panels (a)--(c) correspond to hybridization amplitudes $\Delta(x=0) = 0.01$,
$0.1$, and $0.4\,\mathrm{eV}$, respectively. White curves are contours at
successive decades of $\mathcal{K}$; all panels share the same color scale and
contour levels.}
    \label{fig:ErpenbeckThoss-Gamma-contour}
\end{figure}

\begin{figure}[H]
    \centering
    \includegraphics[width=1.0\linewidth]{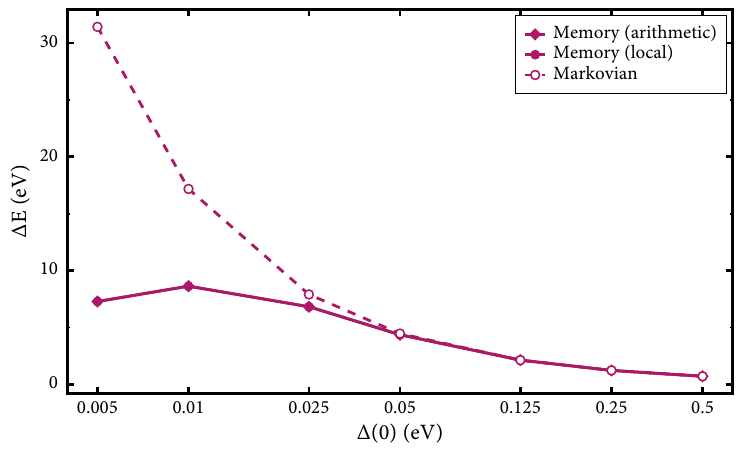}
    \caption{\textbf{Hybridisation dependence of the CPA energy loss in the
    Erpenbeck--Thoss model.} Energy loss $\Delta E$ as a function of the
    hybridisation magnitude $\Delta(0)$ at temperature $T = 300$~K
    and incident translational energy $E_t = 2.0$~eV. Three CPA friction schemes
    are compared: Markovian (open circles, dashed line), and memory friction with
    arithmetic (filled diamonds) and local (filled circles) kernel averaging.}
    \label{fig:ErpenbeckThoss_DeltaE_Gamma_SM}
\end{figure}

\begin{figure}[H]
    \centering
    \includegraphics[width=1.0\linewidth]{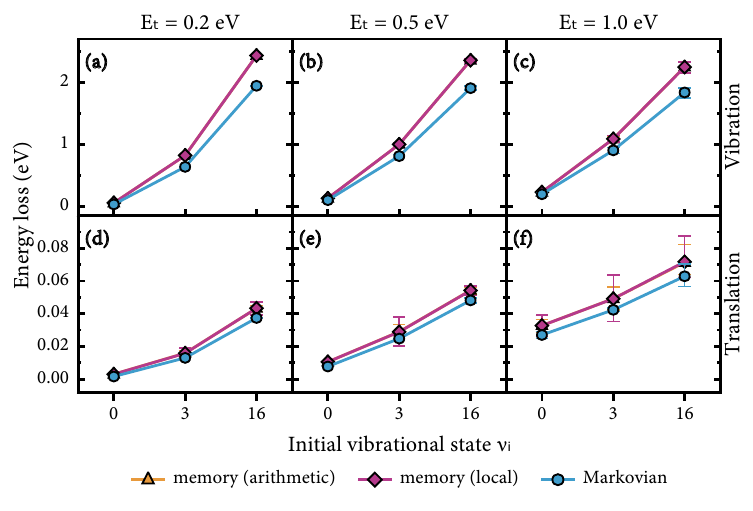}
    \caption{\textbf{Vibrational (top row, \textbf{a}--\textbf{c}) and translational (bottom row,
\textbf{d}--\textbf{f}) energy loss of the NO/Au(111) NAH system at temperature $T = 300$~K, obtained from CPA at three incident translational energies
$E_t = 0.2$, $0.5$ and $1.0$~eV.} Three friction schemes are compared:
Markovian friction, and memory friction with arithmetic and local
kernel averaging. Each marker demontrates the ensemble mean over 1000 trajectories; error bars span the 2.5--97.5 percentile interval of the
per-trajectory energy-loss distribution.}

    \label{fig:GHM_virbation_translation_loss}
\end{figure}

\begin{figure}[H]
    \centering
    \includegraphics[width=1.0\linewidth]{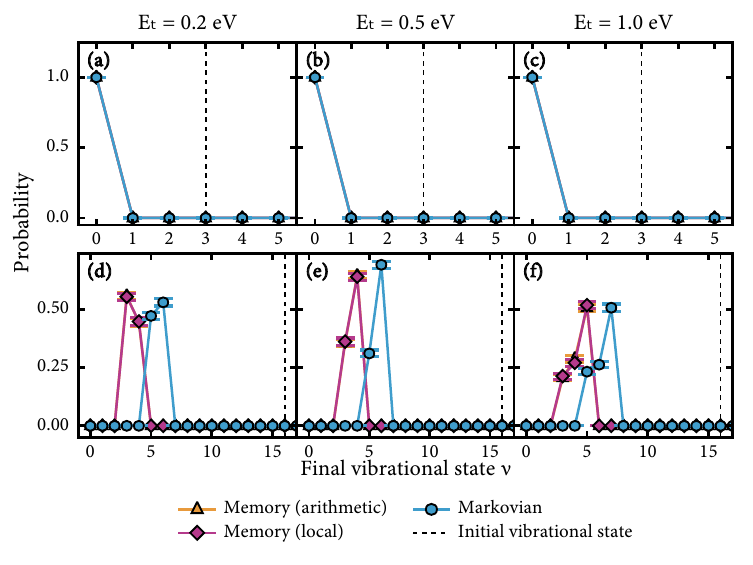}
\caption{\textbf{State-resolved vibrational relaxation of NO scattered from
Au(111).} Final vibrational-state distributions $P(v_\text{f})$ at $T = 300$~K for
initial states $v_i = 3$ (top row) and $v_i = 16$ (bottom row) and incident
translational energies $E_t = 0.2$, $0.5$, and $1.0$~eV (columns). Vibrational quantum states have been determined with the Einstein-Brillouin-Keller method. Trajectories where energies nominally fall into a negative vibrational quantum ($v<0$) are set to $v=0$. Three CPA
friction schemes are compared: Markovian, and memory friction with the local (red) and arithmetic (yellow) kernel averaging. Markers give $P(v_\text{f})$ from 1000 trajectories each; error bars are the binomial standard error, and dashed vertical lines
mark $v_i$.}

    \label{fig:GardnerMaurerCPA_SM}
\end{figure}

\begin{figure}[H]
    \centering
    \includegraphics[width=1.0\linewidth]{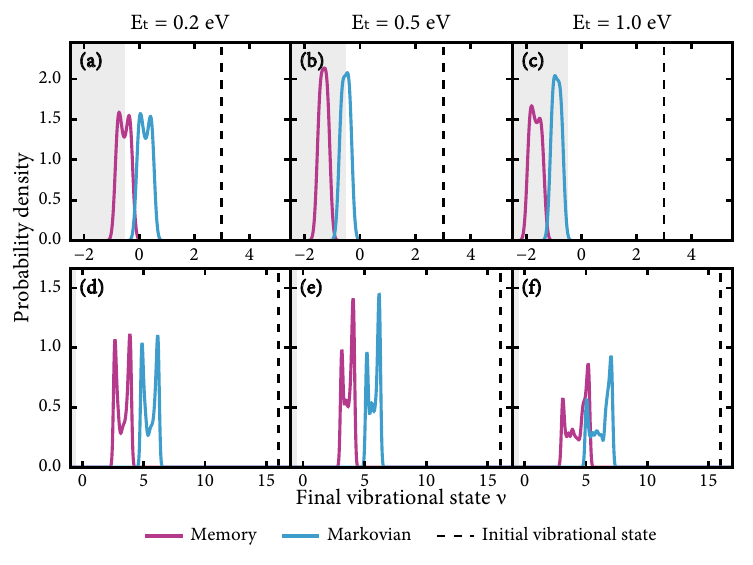}
\caption{\textbf{Continuous state-resolved vibrational relaxation of NO scattered from
Au(111).} Final vibrational-state probability density $P(v_\text{f})$ at $T = 300$~K for
initial states $v_i = 3$ (top row) and $v_i = 16$ (bottom row) and incident
translational energies $E_t = 0.2$, $0.5$, and $1.0$~eV (columns). Vibrational states are determined with Einstein-Brillouin-Keller analysis. Due to the CPA, negative vibrational frequencies can arise, as the trajectory does not respond to energy loss. Two CPA friction
schemes are compared: Markovian (blue) and memory friction with local kernel averaging
(magenta). Each curve is a Gaussian kernel density estimate, with width $\sigma = 0.1$
and normalised to unit area, of the continuous EBK quantum numbers $v_\text{f}$ of 1000
trajectories. Dashed vertical lines mark $v_i$. The grey shading marks
$v_\text{f} < -\tfrac{1}{2}$, the region below the bottom of the NO binding potential:
$v_\text{f} = -\tfrac{1}{2}$ corresponds to the potential minimum $V_{\min}$, so
trajectories in the shaded region have lost more energy to friction than the NO
vibration held above $V_{\min}$.}

    \label{fig:GardnerMaurerContinuousCPA_SM}
\end{figure}

\section{Negativity analysis of frequency-dependent electronic friction for NO/Au(111) NAH system}
As shown in Figs.~3(d) and 3(h) of the main text, the smallest eigenvalue of the
frequency-dependent electronic friction tensor, $\lambda_{\min}(\mathbf{x},\omega)$, becomes negative in the NO/Au(111) NAH system (Gardner--Habershon--Maurer model). A negative eigenvalue means that, through the memory kernel, energy can flow back from the Au(111) surface electrons to the NO molecule. To measure the magnitude of this effect, we perform the CPA analysis
within the Markovian approximation, using at each configuration the most negative
eigenvalue over all frequencies, $\tilde{\lambda}(\mathbf{x}) = \min_{\omega}
\lambda_{\min}(\mathbf{x},\omega)$, as the friction along $z$. Evaluated along the
Born--Oppenheimer molecular dynamics trajectory $\mathbf{x}^{\mathrm{BO}}(t) = \big(r^{\mathrm{BO}}(t),
z^{\mathrm{BO}}(t)\big)$, this gives

\begin{equation}
\begin{aligned}
  \Delta E_{\text{neg}}^{f}
  &= \int_{0}^{t_{\text{traj}}} \dot{\mathbf{x}}^{\mathrm{BO}}(t)^{\top}
     \begin{pmatrix} 0 & 0 \\ 0 & \tilde{\lambda}\big(\mathbf{x}^{\mathrm{BO}}(t)\big) \end{pmatrix}
     \dot{\mathbf{x}}^{\mathrm{BO}}(t)\,\mathrm{d}t \\
  &= \int_{0}^{T} \tilde{\lambda}\big(\mathbf{x}^{\mathrm{BO}}(t)\big)\,
     \big[\dot{z}^{\mathrm{BO}}(t)\big]^{2}\,\mathrm{d}t \;\le\; 0,
\end{aligned}
\end{equation}
where $t_{\text{traj}}$ is the duration of the trajectory. Because $\tilde{\lambda}$ is the most
negative friction the molecule can experience at each configuration,
$|\Delta E_{\text{neg}}^{f}|$ is an upper bound to the energy returned to the NO molecule by the
negative part of the friction spectrum of a single trajectory. As shown in Figure S6, this leads to a total energy gain of at most 0.2\% of the total energy loss during dynamics, which will not significantly affect the energy loss or dynamics.

\begin{figure}[H]
    \centering
    \includegraphics[width=1.0\linewidth]{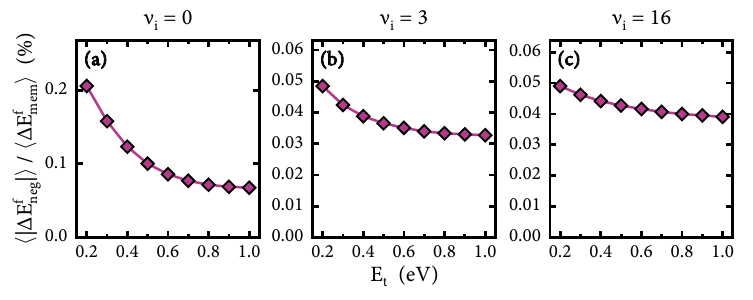}
    \caption{\textbf{Upper bound on the energy returned by negative friction eigenvalues in the NO/Au(111) NAH system at $T = 300$~K.}
Ratio $\langle|\Delta E^{f}_{\mathrm{neg}}|\rangle / \langle\Delta E^{f}_{\mathrm{mem}}\rangle$ as a function of incident translational energy $E_t$ for initial vibrational states (a) $v_i = 0$, (b) $v_i = 3$ and (c) $v_i = 16$.
$\Delta E^{f}_{\mathrm{mem}}$ is the CPA energy loss with memory friction and local kernel averaging.
Averages are over 1000 trajectories.
The ratio stays below $0.21\,\%$ for all $E_t$ and $v_i$.}
\label{fig:negative_bound_ratio}

\end{figure}

\section{Time scale analysis of memory-dependent electronic friction for Erpenbeck--Thoss model}
\begin{figure}[H]
    \centering
    \includegraphics[width=1.0\linewidth]{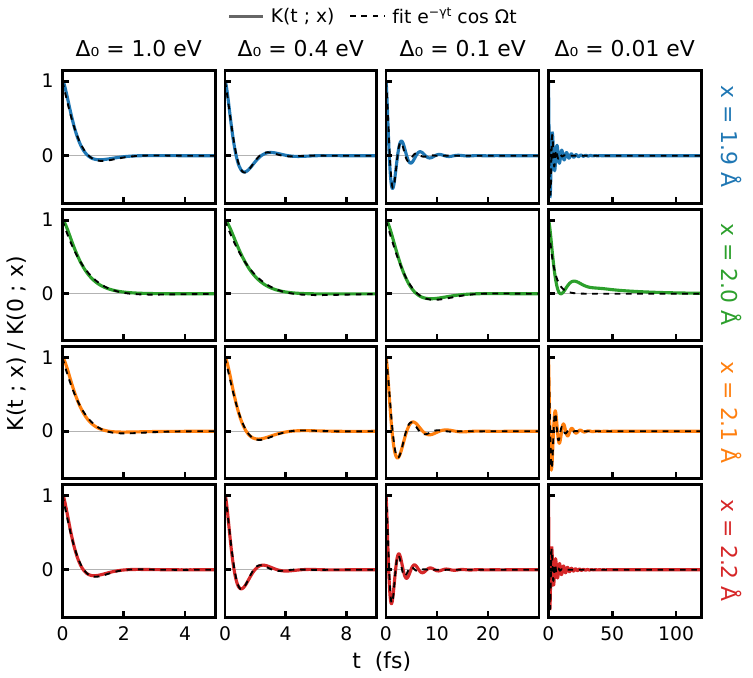}
\caption{%
  \textbf{Damped-cosine fits to the time-domain memory kernel of the Erpenbeck--Thoss
  model at $T = 300$~K.} Normalised kernels $K(t;x)/K(0;x)$ (solid, coloured by nuclear
  position, one row per $x = 1.9$--$2.2$~\AA) and least-squares fits of
  $e^{-\gamma t}\cos\Omega t$ (black dashed). Columns (left to right) show coupling
  $\Delta_0 = 1.0$, $0.4$, $0.1$ and $0.01$~eV, with time windows of 5, 10, 30 and
  120~fs, respectively; each curve is fitted over its whole window. The fitted decay
  rates $\gamma$ are listed in \cref{tab:ET_kernel_fit_SM}. The fits capture the decay of
  every kernel except at $x = 2.0$~\AA{} and $\Delta_0 = 0.01$~eV, where the fit reduces
  to a single exponential ($\Omega = 0$) and misses the bump near 20~fs and the slowly
  decaying tail.
  \label{fig:ET_kernel_fit_SM}
}

\end{figure}

\begin{table}[htbp]
\caption{\textbf{Memory times of the time-domain memory kernel in the Erpenbeck--Thoss model.}
Memory time $\tau = 1/\gamma$, where $\gamma$ is the decay rate from least-squares fits of
$e^{-\gamma t}\cos\Omega t$ to the normalised kernels $K(t;x)/K(0;x)$ at $T = 300$~K
(\cref{fig:ET_kernel_fit_SM}); $\Omega$ only allows the fit to follow the oscillations of
the kernel. After $\tau$ the envelope $e^{-\gamma t}$ has fallen to $1/e$ of its initial
value. At $\Delta_0 = 0.01$~eV and $x = 2.0$~\AA{}, next to the level crossing
$x_c = 2.02$~\AA{}, the fit returns $\Omega = 0$ and does not capture the slow tail of the
kernel, so $\tau$ underestimates its memory there. All times in
\si{\femto\second}.\label{tab:ET_kernel_fit_SM}}
\begin{ruledtabular}
\begin{tabular}{S[table-format=1.2]S[table-format=1.2]S[table-format=1.2]S[table-format=1.2]S[table-format=1.2]}
{$\Delta_0$ (\si{\eV})} & {$x = 1.9$~\AA} & {$x = 2.0$~\AA} & {$x = 2.1$~\AA} & {$x = 2.2$~\AA} \\
\hline
1.00 & 0.56 & 0.79 & 0.67 & 0.52 \\
0.40 & 0.94 & 1.81 & 1.19 & 0.87 \\
0.10 & 1.73 & 4.33 & 2.43 & 1.51 \\
0.01 & 2.39 & 4.22 & 3.83 & 2.00 \\
\end{tabular}
\end{ruledtabular}
\end{table}

\begin{figure}[H]
    \centering
    \includegraphics[width=1.0\linewidth]{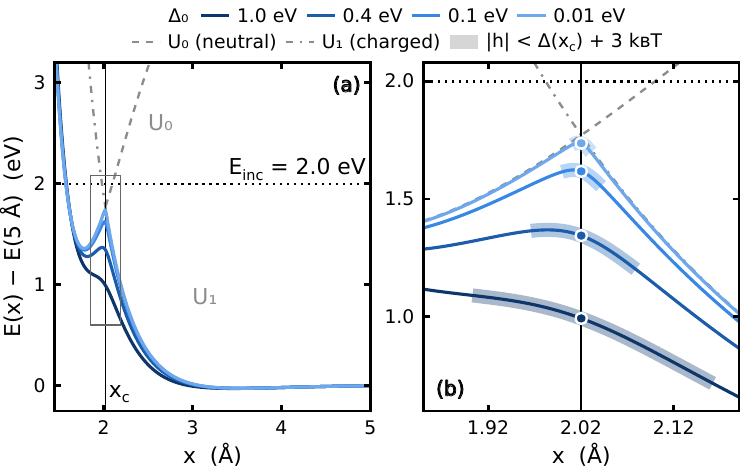}
\caption{\textbf{Ground-state potential energy surfaces for H scattering in the
Erpenbeck--Thoss model.} (a) Mean-field ground-state energy
relative to $E(5\,\text{\AA})$,
on which a hydrogen atom moves, for hybridisation strengths $\Delta_0 = 1.0$, $0.4$,
$0.1$ and $0.01$~eV (dark to light blue) at $T = 300$~K. The diabatic curves $U_0$ (neutral, dashed) and $U_1$
(charged, dash-dotted) are shown on the same energy reference. $x_c$ (vertical line) represents the crossing point where $U_1(x_c) = U_0(x_c)$. 
$\Delta_0$ only sets how sharp the switch is. The dotted line marks the incident
energy $E_\text{inc} = 2.0$~eV, and the grey box marks the region enlarged in (b).
(b) Enlargement around $x_c$. Dots mark the barrier $E(x_c)$; the gap between each dot
and $E_\text{inc}$ is the atom's kinetic energy at $x_c$. Shaded bands mark the friction
window $|h(x)| < \Delta(x_c) + 3k_\text{B}T$, where the impurity level lies close
enough to the Fermi level for electronic friction to be significant.}
\label{fig:ET_H_scattering_PES_SM}

\end{figure}

\begin{table}[htbp]
\caption{\textbf{Classical H-atom scattering trajectories in the Erpenbeck--Thoss model.}
One trajectory per hybridisation strength $\Delta_0$: an H atom starts at $x = 5$~\AA{}
moving towards the surface with an incident energy of 2.0~eV and moves on the mean-field
surface $E(x)$ of \cref{fig:ET_H_scattering_PES_SM}(a) at $T = 300$~K, without electronic
friction. Scattering time: time from leaving $x = 5$~\AA{} to returning there. Turning
point: closest approach to the surface, and the time at which it is reached. Crossings of
$x_c$: times at which the atom passes the crossing point $x_c = 2.020$~\AA{}, where
$h(x_c) = 0$, on the way in and on the way out.\label{tab:ET_H_trajectories_SM}}
\begin{ruledtabular}
\begin{tabular}{S[table-format=1.3]S[table-format=2.1]S[table-format=1.3]S[table-format=2.1]c}
{$\Delta_0$ (\si{\eV})} & {Scattering time (\si{\femto\second})} & {Turning point (\si{\angstrom})}
  & {Turning time (\si{\femto\second})} & {Crossings of $x_c$ (\si{\femto\second})} \\
\hline
1.000 & 40.2 & 1.582 & 20.1 & 15.7, 24.5 \\
0.400 & 41.6 & 1.578 & 20.8 & 16.0, 25.6 \\
0.100 & 42.8 & 1.576 & 21.4 & 16.2, 26.5 \\
0.010 & 43.4 & 1.576 & 21.7 & 16.4, 27.0 \\
\end{tabular}
\end{ruledtabular}
\end{table}

\begin{table}[htbp]
\caption{\textbf{Properties at the crossing point $x_c$ of the trajectories in
\cref{tab:ET_H_trajectories_SM}.} Arrival: time at which the atom first reaches $x_c$.
Barrier: mean-field energy $E(x_c)$ relative to $x = 5$~\AA{} (dots in
\cref{fig:ET_H_scattering_PES_SM}(b)). Residence time: time per pass that the atom spends
inside the friction window $|h(x)| < \Delta(x_c) + 3k_\text{B}T$ (shaded bands in
\cref{fig:ET_H_scattering_PES_SM}(b)); it is the same on the way in and on the way
out.\label{tab:ET_H_crossing_SM}}
\begin{ruledtabular}
\begin{tabular}{S[table-format=1.3]S[table-format=2.1]S[table-format=1.3]S[table-format=1.2]}
{$\Delta_0$ (\si{\eV})} & {Arrival (\si{\femto\second})} & {Barrier (\si{\eV})}
  & {Residence time (\si{\femto\second})} \\
\hline
1.000 & 15.7 & 0.993 & 1.84 \\
0.400 & 16.0 & 1.345 & 1.00 \\
0.100 & 16.2 & 1.618 & 0.49 \\
0.010 & 16.4 & 1.737 & 0.29 \\
\end{tabular}
\end{ruledtabular}
\end{table}

\section{Time scale analysis of memory-dependent electronic friction for Gardner--Habershon--Maurer model}

\begin{figure}[H]
    \centering
    \includegraphics[width=1.0\linewidth]{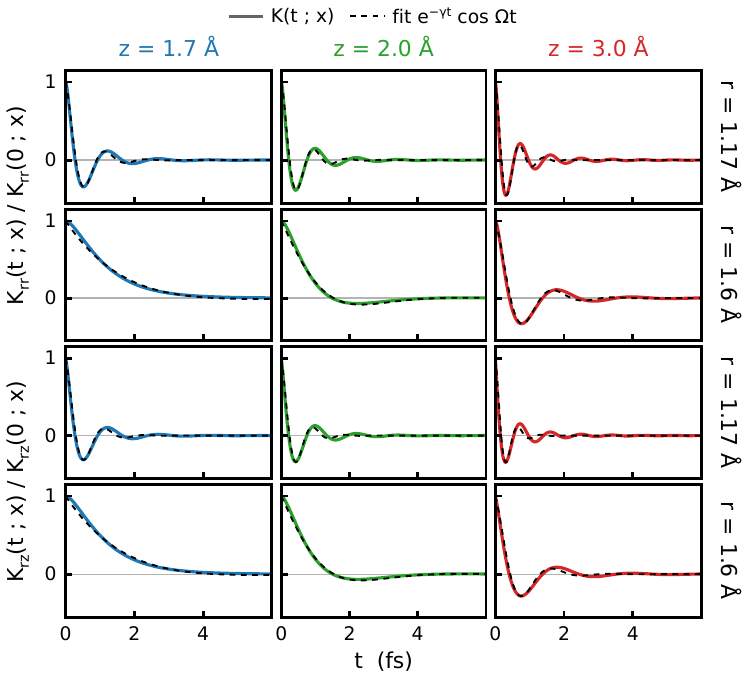}
\caption{%
  \textbf{Damped-cosine fits to the time-domain memory kernel of NO on Au(111) at
  $T = 300$~K.} Normalised kernel elements $K_{rr}(t;x)/K_{rr}(0;x)$ (top two rows) and
  $K_{rz}(t;x)/K_{rz}(0;x)$ (bottom two rows) of the GHM model at N--O bond lengths
  $r = 1.17$ and $1.6$~\AA{} (solid, coloured by molecule--surface distance
  $z = 1.7$, $2.0$ and $3.0$~\AA{}, one column each), with least-squares fits of
  $e^{-\gamma t}\cos\Omega t$ over 0--6~fs (black dashed). The fitted decay rates
  $\gamma$ are listed in \cref{tab:NOAu_kernel_fit_SM}. At $r = 1.17$~\AA{} the fits
  follow the first oscillation, but the later, weaker oscillations decay more slowly
  than the fitted envelope.
  \label{fig:NOAu_kernel_fit_SM}
}

\end{figure}

\begin{table}[htbp]
\caption{\textbf{Memory times of the time-domain memory kernel in the
Gardner--Habershon--Maurer (GHM) model of NO on Au(111).}
Memory time $\tau = 1/\gamma$, where $\gamma$ is the decay rate from least-squares fits of
$e^{-\gamma t}\cos\Omega t$ to the normalised elements $K_{rr}(t;x)/K_{rr}(0;x)$ and
$K_{rz}(t;x)/K_{rz}(0;x)$ at $T = 300$~K, for N--O bond length $r$ and
molecule--surface distance $z$ (\cref{fig:NOAu_kernel_fit_SM}); $\Omega$ only allows the
fit to follow the oscillations of the kernel. All times in
\si{\femto\second}.\label{tab:NOAu_kernel_fit_SM}}
\begin{ruledtabular}
\begin{tabular}{lS[table-format=1.2]S[table-format=1.2]S[table-format=1.2]S[table-format=1.2]}
Element & {$r$ (\si{\angstrom})} & {$z = 1.7$~\AA} & {$z = 2.0$~\AA} & {$z = 3.0$~\AA} \\
\hline
$K_{rr}$ & 1.17 & 0.53 & 0.49 & 0.43 \\
         & 1.60 & 1.54 & 1.08 & 0.76 \\
$K_{rz}$ & 1.17 & 0.48 & 0.42 & 0.30 \\
         & 1.60 & 1.54 & 1.06 & 0.64 \\
\end{tabular}
\end{ruledtabular}
\end{table}

\begin{table}[htbp]
\caption{\textbf{Vibrational parameters of gasphase NO in the GHM model.}
NO starts at the outer turning point $r_{+}$ (most stretched bond) of the EBK vibrational
level $\nu$, for $\nu = 3$ and $16$, the initial states of the scattering simulations. It
starts with zero bond velocity at molecule--surface distance $z = 10$~\AA{}, with no
translational motion, and moves on the adiabatic ground-state surface. $T_{\text{vib}}$ is the period
of one vibration (most stretched back to most stretched), $T_{\text{vib}}/2$ the time from most
stretched to most compressed ($r_{-}$), $\tilde{\nu} = 1/(cT_{\text{vib}})$ the wavenumber and
$\omega_{\text{ang}} = 2\pi/T_{\text{vib}}$ the angular frequency. \label{tab:NO_vib_period_SM}}
\begin{ruledtabular}
\begin{tabular}{S[table-format=2.0]S[table-format=2.3]S[table-format=2.3]S[table-format=1.4]S[table-format=1.4]S[table-format=4.0]S[table-format=1.4]}
{$\nu$} & {$T_{\text{vib}}$ (\si{\femto\second})} & {$T_{\text{vib}} /2$ (\si{\femto\second})} & {$r_{+}$ (\si{\angstrom})}
  & {$r_{-}$ (\si{\angstrom})} & {$\tilde{\nu}$ (\si{\per\centi\metre})} & {$\omega_{\text{ang}}$ (\si{\per\femto\second})} \\
\hline
3  & 18.511 & 9.255  & 1.3079 & 1.0446 & 1802 & 0.3394 \\
16 & 24.841 & 12.420 & 1.6031 & 0.9596 & 1343 & 0.2529 \\
\end{tabular}
\end{ruledtabular}
\end{table}

\section{Overview of trajectories analysed with DFT electronic friction}

\begin{table}[htb]
\centering
\caption{\textbf{Overview of the six $v_i=16$ BOMD \ce{NO}/Au(111) trajectories used in the memory-friction analysis.} The final vibrational action $N_f$ and rotational angular momentum $J_f$ are taken from the VENUS product analysis.}
\label{tab:v16-bomd-trajectories}
\begin{tabular}{cccc}
\toprule
Trajectory & $N_f$ & $J_f$ & Length / fs \\
\midrule
1  & 4.88 & 40.47 & 383.9 \\
2  & 6.31 & 65.95 & 410.0 \\
3 & 5.92 & 61.49 & 400.0  \\
4 & 4.14 & 79.02 & 394.9 \\
5 & 5.95 & 68.82 & 400.0 \\
6 & 8.98 & 17.00 & 450.6  \\
\bottomrule
\end{tabular}
\end{table}

\begin{figure}[H]
    \centering
    \includegraphics[width=\linewidth]{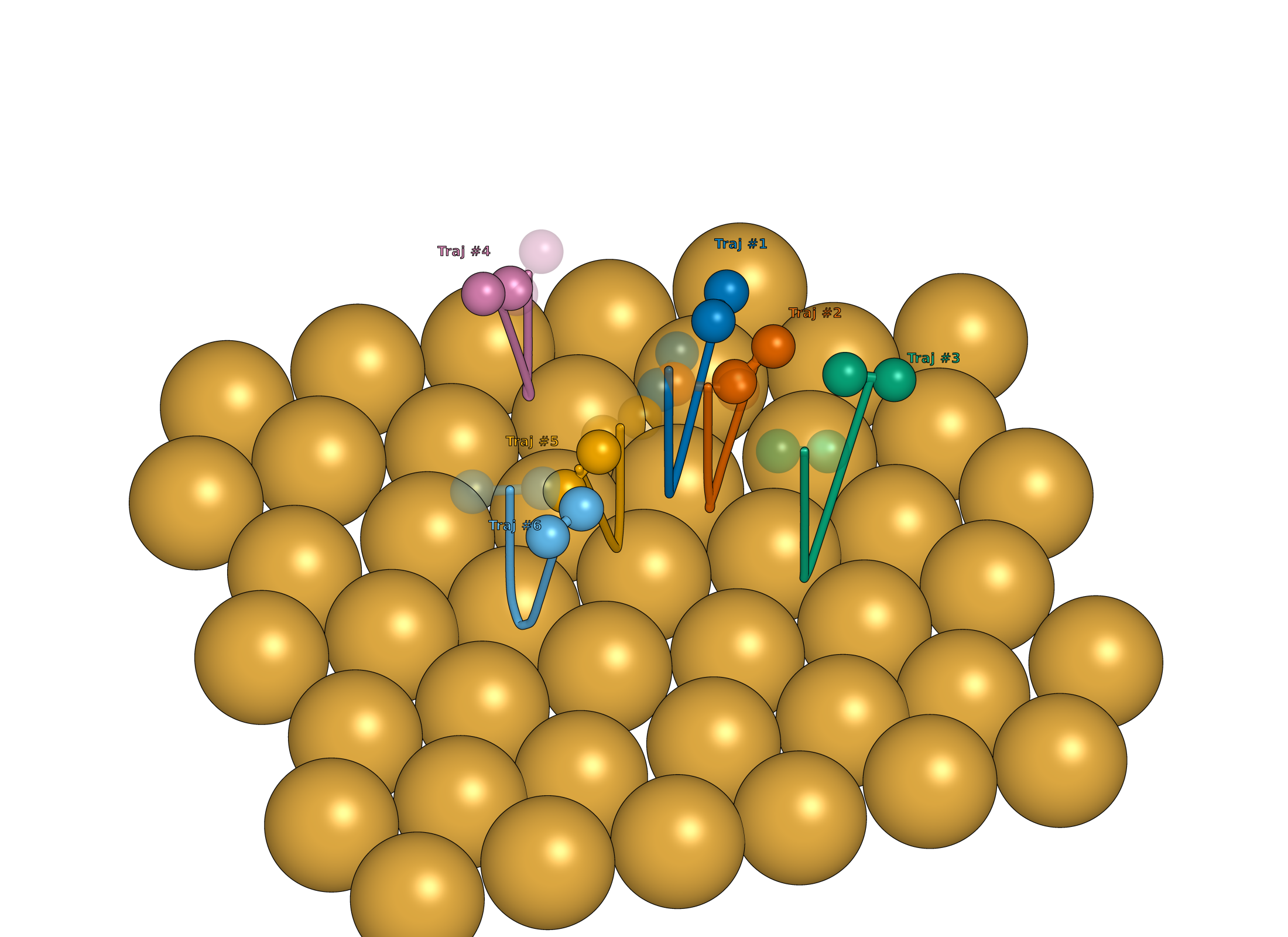}
    \caption{\textbf{Ball and stick model visualising the six classical molecular dynamics trajectories employed for the classical path approximation analysis of energy dissipation.} Transparent atoms indicate the starting configuration of the trajectory, solid atoms indicate the final configuration. Different colours are used for N and O atoms of the six different trajectories.}
    \label{fig:placeholder}
\end{figure}

\section{Numerical stability of analysis for ab initio friction dissipation}

\begin{figure}[H]
    \centering
    \includegraphics[width=0.5\linewidth]{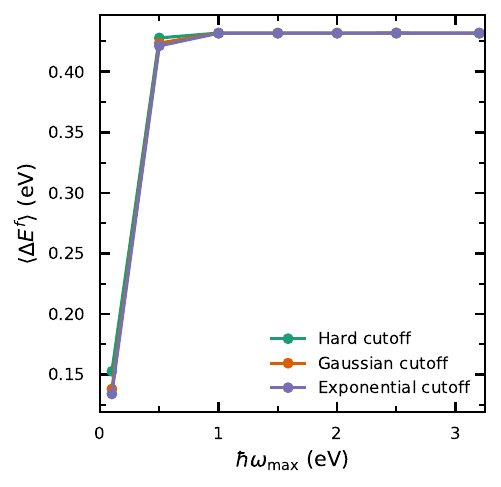}
    \caption{\textbf{Sensitivity of CPA frictional dissipation to the high-energy cutoff.} Mean final frictional energy loss, $\langle \Delta E^{f} \rangle$, averaged over six NO/Au(111) BOMD trajectories, as a function of the maximum excitation energy $\hbar\omega_{\max}$ used in the memory-kernel reconstruction. Results are shown for hard, Gaussian, and exponential cutoff functions applied to the $
    \sigma=0.01$~eV friction spectrum. The memory kernel is constructed with the one-sided cosine transform and evaluated using endpoint treatment. }\label{fig:emax_convergence}
\end{figure}

\begin{figure}
    \centering
    \includegraphics[width=0.5\linewidth]{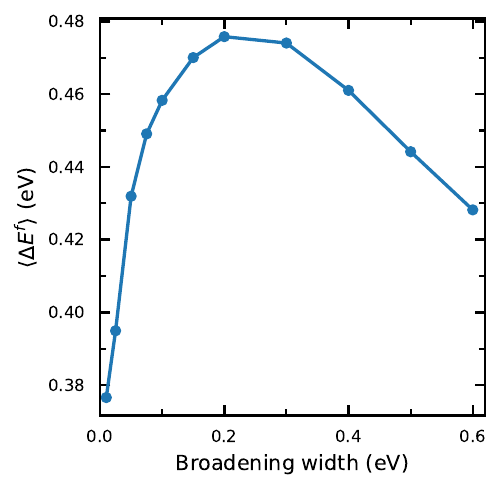}
    \caption{\textbf{Dependence of CPA memory-friction dissipation on spectral broadening.} Mean final frictional energy loss, $\langle \Delta E^{f} \rangle$, averaged over six NO/Au(111) BOMD trajectories, as a function of the Gaussian broadening width used to smooth the positive-frequency friction spectrum before reconstructing the memory kernel. Starting from $\sigma=0.01$ eV spectrum output from FHI-aims, additional Gaussian broadening was applied in quadrature to reach the target total width at each point. The spectrum used $\hbar\omega_\mathrm{max}=3.2$~eV and an exponetial cutoff as in \Cref{fig:emax_convergence}. The memory kernel was then constructed using local endpoint treatment.} \label{fig:broad_convergence}
\end{figure}

\begin{figure}[H]
    \centering
    \includegraphics[width=\linewidth]{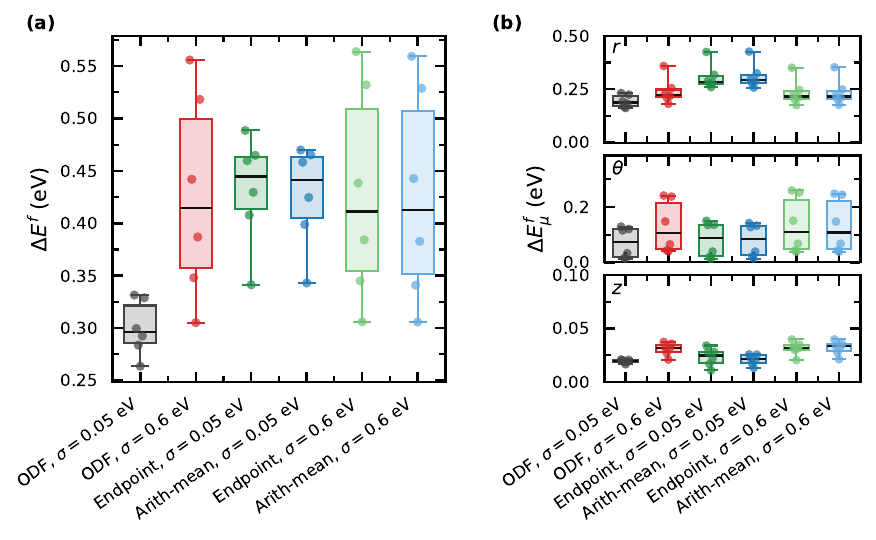}
\caption{\textbf{Sensitivity of CPA frictional dissipation to the choice of local memory-kernel symmetrization and spectral broadening.} Final CPA frictional energy losses, $\Delta E^{f}$, evaluated over six representative NO/Au(111) trajectories for two local memory-kernel constructions: the endpoint kernel and the arithmetic-mean kernel (see main manuscript). Results are shown for memory kernels reconstructed from friction spectra broadened to total widths of $\sigma = 0.05$ and $0.6$~eV. In each case, the time-domain kernel was obtained from the normalized one-sided cosine transform with an exponential smooth cutoff at $\hbar\omega_{\max}=3.2$~eV. Panel (a) shows the distribution of total final frictional energy losses, while panel (b) resolves the corresponding mode contributions into the internal stretch coordinate $r$, the orientation angle $\theta$, and the centre-of-mass motion along $z$. ODF results evaluated with $\sigma=0.05$~eV and $\sigma=0.6$~eV are included to compare
Markovian and memory-dependent dissipation at equal spectral
broadening.}
    \label{fig:abinitio_approxs}
\end{figure}

\bibliography{SM_bibliography}